%% file: main.tex
\documentclass[conference]{IEEEtran}
\IEEEoverridecommandlockouts
\usepackage[utf8]{inputenc}
\usepackage{authblk}

\usepackage{amsmath,amsfonts,amssymb,amsthm}
\usepackage{newtxmath}
\usepackage{thmtools, thm-restate}
\usepackage{nicefrac}
\usepackage{cite}
\usepackage{enumerate}
\usepackage{subcaption}
\usepackage{makecell}
\usepackage{hyperref} 

\usepackage{tikz}
\usetikzlibrary{arrows, positioning, chains}

\usepackage{tabularx,booktabs}
\newcolumntype{Y}{>{\centering\arraybackslash}X}

\usepackage{multicol}

\usepackage[skins]{tcolorbox}
\tcbuselibrary{breakable}

\newcommand{\citep}[1]{\cite{#1}}

\newtheorem{theorem}{Theorem}
\newtheorem{corollary}[theorem]{Corollary}
\newtheorem{lemma}[theorem]{Lemma}

\newtheorem{definition}{Definition}

\newtheorem{remark}{Remark}

\newenvironment{inproof}{\begin{proof}}{\end{proof}}

\definecolor{draw_orange}{rgb}{0.92, 0.57, 0.05} 

\input{MacrosMaths}

\author[1,2,3]{Natasha Fernandes}
\author[1]{Annabelle McIver}
\author[2]{Catuscia Palamidessi}
\author[3]{Ming Ding}
\affil[1]{Macquarie University, Sydney, Australia}
\affil[2]{Institut Polytechnique de Paris, France}
\affil[3]{Data61, Sydney, Australia}

\begin{document}

\title{Universal Optimality and Robust Utility Bounds for Metric Differential Privacy}


\maketitle

\begin{abstract}


We study the privacy-utility trade-off in the context of metric differential privacy. Ghosh et al.   introduced the idea of \emph{universal optimality} to characterise
the ``best'' mechanism for a certain query that simultaneously satisfies (a fixed) $\epsilon$-differential privacy constraint whilst at the same time providing better utility compared to any other $\epsilon$-differentially private mechanism for the same query.  They showed 
that the Geometric mechanism is ``universally optimal'' for the class of ``counting queries''. On the other hand, Brenner and Nissim  showed that outside the space of counting queries, 
and for the specific loss function called $\lbin$, 
no such universally optimal mechanisms exist.
Except for universal optimality of the Laplace mechanism, there have been no generalisations of these universally optimal results to other classes of differentially-private mechanisms. 

In this paper we use metric differential privacy and quantitative information flow as the fundamental principle for studying universal optimality.  
Metric differential privacy is a generalisation of both 
standard (i.e., central) differential privacy and local differential privacy, and it is increasingly being used in various application domains, for instance in location privacy and in privacy preserving machine learning.  As do Ghosh et al. and Brenner and Nissim, we measure utility in terms of  loss functions, and we interpret the notion of a privacy mechanism as an information-theoretic channel satisfying constraints defined by $\epsilon$-differential privacy and a metric meaningful to the underlying state space. Using this framework we are able to clarify Nissim and Brenner's negative results, showing (a) that in fact \emph{all} privacy types 
contain optimal mechanisms relative to certain kinds of non-trivial loss functions, and (b) extending and generalising their negative results beyond $\lbin$ specifically to a wide class of non-trivial loss functions.

Our exploration suggests that universally optimal mechanisms are indeed rare within privacy types. We therefore propose weaker universal benchmarks of utility called \emph{privacy type capacities}.  We show that such capacities always exist and  can be computed using a convex optimisation algorithm.

We illustrate these ideas on a selection of examples with several different underlying metrics.
\end{abstract}


\section{Introduction}

One of the main 
challenges in privacy is to devise mechanisms that protect the   sensitive information contained in the data, while at the same time preserving an acceptable degree of utility. 
Differential privacy~\cite{Dwork:06:TCC,Dwork:06:ICALP}, one of the most successful frameworks for privacy protection, allows access to a dataset only via querying the dataset through an interface controlled by the data curator, 
which ensures privacy 
by adding controlled random noise to the result of the query before reporting it. In this setting, the trade-off with utility is determined by how the noise is calibrated: a ``bad'' calibration may cause a lot of utility loss without necessarily improving  privacy. Of course, the ideal   mechanisms are the Pareto optimal ones, i.e., those that offer the best utility for a given level of privacy.

There are many notion of utility, depending on the goals and the means of the data consumers. The goals are usually formalised in terms of \emph{loss functions}, and an  important class of consumers is the so-called \emph{Bayesian consumers}, who may have some prior knowledge about the data and are able to make the most out of the reported answer, in the sense of deriving (via Bayesian inference, using their prior)  the information that minimises the  loss with respect to the true answer.  

In the context of the Bayesian notion of utility, an important desideratum for a mechanism is to be optimal for all consumers, no matter their prior. This is in line with the property of differential privacy, which does not depend on the prior knowledge of the attacker. 
This independence from the prior is appealing because when we design mechanisms we don't know what kind of knowledge attackers and consumers may acquire over time.

The goal of this paper is study universally optimal mechanisms  for Bayesian consumers over any class of loss functions (ie. not restricted to the monotone class). While monotone loss functions capture many consumers of interest, other research~\cite{alvim2019science} provides examples of non-monotonic consumers which we argue should be included in a complete study of optimality. Moreover, the non-monotone $\lnib$ loss function (introduced later in this paper) is crucial in computing capacities for classes of privacy mechanisms.

To better understand the issues that arise in this context, let us introduce some notation. 

Let ${\cal C}$ be a class of (Bayesian) consumers, where each consumer is identified by a prior $\pi$ and a loss function $\ell$, and let ${\cal D}$ be a class of $\epsilon$-differentially private mechanisms. 
A mechanism $M \in {\cal D}$ probabilistically transforms the true answer (input) into a reported answer (output, observation)\footnote{This functionality of the mechanism corresponds to the so-called ``oblivious'' model, in which the noise depends only on the result of the query and not on the original dataset.}, and it is called
\emph{universally optimal} w.r.t.\ ${\cal C}$ 
if $M$ provides the best utility for every consumer in  ${\cal C}$ compared with any other $M'\in {\cal D}$. ``Best utility'' for a consumer w.r.t. a mechanism $M$ is the minimal value of the consumer's loss function over all possible remappings from observations to inputs, i.e.,
\begin{equation}\label{eq:utility_loss}
  U_\ell(\pi, M) \Wide{=} \min_r \sum\limits_{x \In X} \pi_x \sum\limits_{y \In Y} M_{x,y} \ell(r(y), x) 
\end{equation}
for inputs $X$, observations $Y$ and remapping function $r$. The universally optimal mechanism $M$ is one for which $U_\ell(\pi, M) \leq U_\ell(\pi, M')$ for all other mechanisms $M' \in {\cal D}$, priors $\pi$ and all loss functions $\ell$ within the class  ${\cal C}$.

Ghosh et al.\ \cite{ghosh2012universally} found that a universally optimal mechanism exists for a particular class ${\cal D}$ of ``counting query'' mechanisms and the class ${\cal C}$ of ``monotonic'' consumers, i.e., described by a monotonic loss function. 
Conversely, 
Brenner and Nissim \cite{brenner2010impossibility} showed that no such optimal mechanism exists for the same class of consumers with respect to  various queries. 
Moreover, they 
proved  that for the specific loss function 
$\lbin$ such optimal mechanisms never exist outside the linear structures of inputs corresponding to the counting queries.

The above 
results are in the context of the so-called \emph{central model} of differential privacy, which assumes that the noise is added by a trusted curator. In recent years, however, the so-called \emph{local model}~\cite{Duchi:13:FOCS} has become more and more popular, also due to the interest of large companies such as Google and Apple. In the local model, the noise is added 
directly by the  data producer, so the microdata are already obfuscated. Consequently, this model is more robust w.r.t. potential security breaches,  and  there is no need for a trusted third party.

Recently it has been shown that both the central and local models 
can be unified under a generalised definition of differential privacy known as $\dx$-privacy~\cite{chatzikokolakis2013broadening}, 
which describes a \emph{metric privacy type}.
$\dx$-privacy is increasingly being used in various application domains, for instance in location privacy, where it is known under the name of \emph{geo-indistinguishability}~\cite{andres2013geo}, and in privacy preserving machine learning \cite{DBLP:conf/post/FernandesDM19}.  

\begin{definition}[Metric privacy type]\label{d1651}
Given a metric space $(\CalX, \dx)$, 
we say that a mechanism $M: \CalX{\to}\Dist\CalY$ satisfies $\dx$-privacy
if for all $x, x' \in \CalX$, 
the following constraint holds:
\begin{equation}\label{e1309}
M(x)(Y) \Wide{\leq} e^{\dx(x,x')}M(x')(Y)~,
\end{equation}
where we denote $M(x)(Y)$ the probability that $M$'s output is contained in the subset $Y\subseteq \CalY$. 
We call the class of $\dx$-private mechanisms on $\calx$ the \emph{metric privacy type} associated to $(\CalX, \dx)$ which we denote by $\PrivType$.
\end{definition}

It is easy to see that $\dx$-privacy subsumes both the central and the local model. For example, the metric privacy type for ``counting queries''
(in the central model) is  $\MetricPrivType{\mathbb{N}, \Euclid}$ where $\Euclid$ is the Euclidean distance and $\mathbb{N}$ the natural numbers\footnote{$M$ can also be seen as a mechanism from datasets to reported answers, as is customary in central differential privacy. In this case, $\dx$ would be the Hamming distance.}.  
For local differential privacy it is $\MetricPrivType{\calx, \Discrete}$ for any inputs $\calx$, where $\Discrete$ is the Discrete metric assigning distance $1$ to all $x \neq x' \in {\cal X}$.
Also note that $\dx$-privacy is strictly more general than local differential privacy, as it is able to express the distance between inputs and tune the noise accordingly. This is important, for instance, in machine learning applications, where $\dx$-privacy can privatise the  gradient descent step without the need for clipping.

\subsection{Contributions}\label{sect:contrib}

Our contributions are as follows:

\begin{enumerate}
   \item Using principles from Quantitative Information Flow, we study optimality for metric differential privacy in full generality; that is, $\dx$-private mechanisms which provide the best utility over arbitrary classes of loss functions. This extends the study by Ghosh et al.~\citep{ghosh2012universally} and Brenner and Nissim~\citep{brenner2010impossibility} which considered optimality only for the central model and only for the class of monotonic loss functions.
    \item We provide a complete characterisation of optimality within a privacy type for finite $\CalX$. This characterisation is independent of the workflow (oblivious, local or otherwise) in which the privacy type arises. This result is significant in that it provides a finite class of mechanisms which characterise the space.
    \item We show that every privacy type contains universally optimal mechanisms (for non-trivial loss functions). This indicates that the impossibility result of Brenner and Nissim is too strong; we provide a tightening of their result, clarifying that the impossibility result for ``sum queries'' holds for a class of \emph{strictly} monotonic consumers.
    \item We extend the qualitative study of optimality to a quantitative study, finding robust capacity bounds for universally optimal mechanisms within a privacy type. These capacity results provide a tight quantitative bound on the maximum leakage -- a measure of the usefulness of a mechanism within a privacy type. In practice for example, such bounds provide benchmarks for experimental evaluation of implemented mechanisms.
\end{enumerate}

\subsection{Related Work}
The problem of universal optimality for arbitrary privacy types, as modelled by $\dx$-privacy, has not yet been studied in full generality.
Chatzikokolakis et al.~\citep{chatzikokolakis2013broadening} showed that optimal mechanisms for ``sum queries'' can be constructed in the oblivious setting by considering a Manhattan metric, rather than a Hamming metric, on the original datasets.
Recent work by Fernandes et al.~\citep{fernandes2021laplace} considers universal optimality of the Laplace mechanism in the continuous setting, using the same QIF framework as in our paper, although our work focuses on a more general problem of universal optimality restricted to discrete spaces.
Other optimality results have been found in non-Bayesian settings. Gupte et al.~\citep{gupte2010universally} proved a universal optimality result for a type of risk-averse consumer using a \emph{minimax} formulation. Aharya et al.~\cite{acharya2020context} prove an optimality result for binary mechanisms similar to our result of \Cor{cor:1123}, but the focus of their work is on minimax consumers.  Kairouz et al.~\citep{kairouz2016extremal} proved optimality for ``staircase mechanisms'' using $f$-divergence as a utility measure, while Koufogiannis et al. \citep{koufogiannis2015optimality} showed that the Laplace mechanism is optimal for the mean-squared error. Finally, Asi and Duchi \cite{Asi:20} studied near-instance optimality, designed for particular dataset instances, in contrast with our study of optimality over mechanisms. 
It is important to note that the notions of utility induced by the above non-Bayesian consumers are simpler than ours, and their trade-off with privacy is more direct. Also, the question of  universality w.r.t. the prior does not arise. Furthermore, most of the results assume heavy restrictions on the noise functions.

\subsection{Organisation of the Paper}

We begin in \Sec{s1106} with a summary of the results needed from Quantitative Information Flow including how to formulate the Universal optimality problem under the same assumption  of ``Bayesian consumers'' as used by Ghosh et al.\ and Brenner and Nissim. In \Sec{s1225}  we introduce privacy types based on a metric $\dx$ and a set of secrets $\calx$; given these ingredients we provide a complete characterisation of the mechanisms of that type. In \Sec{sect:utility} we provide a thorough investigation of utility in terms of loss functions and how there is a close relationship between the complexity of the loss function and the structure of the underlying privacy mechanisms from which we can deduce both when universal properties exist and when they do not depending on the loss function (eg.\ its complexity) and the composition of the privacy mechanisms. In \Sec{s0856} we introduce privacy type capacities and show that they provide a benchmark for computing how much useful information can be extracted from a privacy mechanism. Finally in \Sec{sec:examples} we provide a range of examples that illustrate these ideas.

Where possible we have sketched proofs in the main body of the paper. Full proofs can be found in the appendix.

\section{Quantitative Information Flow for privacy mechanisms}\label{s1106}
\input{qif}

\section{Characterising metric privacy types}\label{sect:characterisation}\label{s1225}
\input{privacy_types}

\section{Utility for Bayesian consumers}\label{sect:utility}
\input{utility}

\subsection{Robustness results}\label{sect:robustness}
\input{robustness}

\subsection{Impossibility results}\label{impossibilities}
\input{impossibility}

\section{Examples of privacy types and their optimality results}\label{sec:examples}
\input{examples}


\section{Conclusion}
We have extended the notion of universal optimality to  $\dx$-privacy. 
Our principal technique is an interpretation of QIF, 
enabling a geometric interpretation of a class of privacy mechanisms and a characterisation of privacy types in terms of refinement. A principal contribution is to clarify the negative results of Brenner and Nissim, extending them beyond $\ell_{bin}$ to other strictly monotonic mechanisms, and removing the connection to $\epsilon$. Our study of the relationship of utility sets provides techniques for finding new optimality results to new domains beyond sum and counting queries, as illustrated by our example of mechanisims over the hamming cube. Finally where the underlying domain is very complex, making it unlikely to find universally optimal results we introduce the notion of privacy type capacity and show how to compute it for both multiplicative and additive capacities. With these results we provide strong benchmarks for leakage measurements in domains where full formal analysis is difficult.

\bibliographystyle{IEEEtran}
\bibliography{optimality}

\appendix 

\input{appendix3}

\end{document}

%% file: MacrosMaths.tex
\usepackage{stmaryrd}   
\usepackage{xifthen}

\newcommand{\lbin}{\ell_\mathrm{bin}}
\newcommand{\lnib}{\ell_\mathrm{nib}}

\newcommand\FN[1] {{\color{red}\footnote{\color{red}{#1}}}}

\newcommand\Section[2][\empty]
	{\ifx#1\empty\section{#2}
	\else\section{#2}\label{#1}\marginpar{\tiny#1}\fi}
\newcommand\SSection[2][\empty]
	{\ifx#1\empty\subsection{#2}
	\else\subsection{#2}\label{#1}\marginpar{\tiny#1}\fi}
\newcommand\SSSection[2][\empty]
	{\ifx#1\empty\subsubsection{#2}
	\else\subsubsection{#2}\label{#1}\marginpar{\tiny#1}\fi}
\newcommand\BE[1] {\begin{equation}\label{#1}\HangLeft[\qquad]{\tiny #1}~}
\newcommand\EE {\end{equation}}

\renewcommand\FN[1] {\relax}
\renewcommand\Section[2][\empty]
	{\section{#2}\ifx#1\else\label{#1}\fi}
\renewcommand\SSection[2][\empty]
	{\subsection{#2}\ifx#1\else\label{#1}\fi}
\renewcommand\SSSection[2][\empty]
	{\subsubsection{#2}\ifx#1\else\label{#1}\fi}
\renewcommand\BE[1] {\begin{equation}\label{#1}}
\renewcommand\EE {\end{equation}}

\newcommand\TN{\ensuremath{T\kern-.1em,\kern-.1emN}}

\newcommand\UIK {\ensuremath{{\kern.1em\cal U}}}
\newcommand\UIT {\ensuremath{{\cal U}_{\kern.07emT}}}

\newcommand\Bag {\ensuremath{\mathbb B}}
\newcommand\DB {\Bag\kern-.03emR}

\newcommand\XX {\ensuremath{{\cal X}}}
\newcommand\YY {\ensuremath{{\cal Y}}}

\newcommand\HangLeft[2][\empty]
	{\ifx#1\empty\makebox[0pt][r]{#2}\else\makebox[0pt][r]{#2#1}\fi}
\newcommand\HangRight[2][\empty]
	{\ifx#1\empty\makebox[0pt][l]{#2}\else\makebox[0pt][l]{#1#2}\fi}

\newcommand\Times {{\times}}
\newcommand\JD[2] {#1\kern.1em{\triangleright}\kern.1em#2}
\newcommand\Hype[1] {[#1]}
\newcommand\JDH[2] {\Hype{#1\kern.1em{\triangleright}\kern.1em#2}}

\newcommand\NStep {$N$\kern-.1em-step}
\newcommand\NStepped {$N$\kern-.1em-stepped}

\newcommand\DX {\kern.1em{\rm d}}

\newcommand\LTVN[1][\empty]{\ensuremath{{\spreadlips{L}}^\varepsilon_{\ifx#1\empty N\else#1\fi}}}

\DeclareMathAlphabet{\mathpzc}{OT1}{pzc}{m}{it}

\makeatletter
\newcommand{\oset}[3][0ex]{%
  \mathrel{\mathop{#3}\limits^{
    \vbox to#1{\kern-2\ex@
    \hbox{$\scriptstyle#2$}\vss}}}}
\makeatother

\newcommand{\MetricPrivType}[1] {{\cal T}_{#1}}
\newcommand{\PrivType}  {\MetricPrivType{\calx,\dx}}

\newcommand\Sec[1] {\S\ref{#1}}

\newcommand\Eqn[1] {(\ref{#1})}
\newcommand\Lem[1] {Lem.~\ref{#1}}
\newcommand\Thm[1] {Thm.~\ref{#1}}
\newcommand\Prop[1] {Prop.~\ref{#1}}
\newcommand\Cor[1] {Cor.~\ref{#1}}
\newcommand\Def[1] {Def.~\ref{#1}}
\newcommand\Tbl[1] {Table~\ref{#1}}
\newcommand\Fig[1] {Figure~\ref{#1}}
\newcommand\App[1] {Appendix \Sec{#1}}

\newcommand\Apply {\mathbin{\raisebox{.1em}{\kern.05em\scriptsize$\vartriangleright$}}}
\newcommand\ApplyR {\mathbin{\raisebox{.1em}{\kern.05em\scriptsize$\langle$}}}

\newcommand\call		{\mathcal{L}}	
\newcommand\calml		{\mathcal{M\kern-.15emL}}	

\newcommand\calx		{\mathcal{X}}
\newcommand\caly		{\mathcal{Y}}

\newcommand\calw		{\mathcal{W}}

\renewcommand\epsilon{\varepsilon}

\newcommand\dx			{\ensuremath{\mathbf{d}}} 

\newcommand\Discrete		{\dx_\textrm{D}}

\newcommand\Euclid		{\dx_2}
\newcommand\Hamming		{\dx_\textrm{H}}

\newcommand\CalL {{\cal L}}

\newcommand\CalX {{\cal X}}
\newcommand\CalY {{\cal Y}}

\newcommand\Defs			{{:=}\,}
\newcommand\Dist			{\mathbb{D}}

\newcommand\Extend[2]		{{#1}\!\!\uparrow\!\! {#2}}

\newcommand\Fun 			{\mathbin{\rightarrow}}

\newcommand\Gain			{\mathbb{G}}

\newcommand\Hyp[2]		{[#1{\Apply}#2]}

\newcommand\Integer		{\mathbb{Z}}

\newcommand\In			{{:}\,} 
           
\newcommand\Loss			{\mathbb{L}}  

\newcommand\Mechs		{\mathcal{M}}  

\newcommand\Real			{\mathbb R}
\newcommand\RealNN		{\Real_{\ge 0}} 

\newcommand\ChanRef     {\sqsubseteq}

\newcommand\Restrict[2]{{#1}\!\!\downarrow\!\! {#2}}

\newcommand\Set[1]                 {\{#1\}}

\newcommand\Uniform		{\upsilon}

\newcommand\Ul		{U_\ell}

\newcommand\VPo		{V}

\newcommand\Vb              {\VPo_1}

\newcommand{\VPost}[2][]{%
     \ifthenelse{\isempty{#1}}{\VPo_g(#2)}{\VPo_g\Hyp{#1}{#2}}%
}
\newcommand\VgPost[2][]{%
     \ifthenelse{\isempty{#1}}{\VPo_g(#2)}{\VPo_g\Hyp{#1}{#2}}%
}

\newcommand\VBayes[2][]{%
     \ifthenelse{\isempty{#1}}{\Vb(#2)}{\Vb\Hyp{#1}{#2}}%
}

\newcommand\Wide[1] 		{~~~#1~~~} 

\newcommand\Supp[1]		{\lceil#1\rceil} 

\newenvironment{Reason}{\begin{tabbing}\hspace{2em}\= \hspace{1cm} \= \kill}
{\end{tabbing}\vspace{-1em}}
\newcommand\Step[2] {#1 \> $\begin{array}[t]{@{}llll}#2\end{array}$ \\}
\newcommand\StepR[3] {#1 \> $\begin{array}[t]{@{}llll}#3\end{array}$
\` {\RF \makebox[0pt][r]{\begin{tabular}[t]{r}``#2''\end{tabular}}} \\}

\newcommand\RF {\small}

%% file: qif.tex

A probabilistic channel $C$ takes an input $x \in \calx$ and outputs an ``observation'' $y \in \caly$ according to a distribution $\Dist{\caly}$. In the discrete case, such channels are $\XX\Times\YY$ matrices $C$ whose row-$x$, column-$y$ element $C_{x,y}$ is the probability that input $x$ will produce output $y$. The $x$-th row $C_{x,-}$ is thus a discrete distribution in $\Dist{\caly}$. We write $\calx{\to}\Dist\caly$ for the channel type.


Given a \emph{prior} distribution $\pi\In\Dist\XX$ on $\XX$,  the channel $C$ can be applied to $\pi$ to create a joint distribution $J$ in $\Dist(\XX\Times \YY)$, written $\JD{\pi}{C}$ and where $J_{x,y}\,{\Defs}\;\pi_x C_{x,y}$. For that $J$, the left-marginal $\sum_y J_{x,y}$  (denoted $J_{x,\Sigma}$) gives the prior $\pi$ again, i.e.\ the probability that the input was $x$ --- thus $\pi_x\,{=}\,J_{x,\Sigma}$.
The right marginal $J_{\Sigma,y}$ is the probability that the output is $y$, given both $\pi$ and $C$.
The $y$-\emph{posterior} distribution on \XX\ is the \emph{conditional} probability that the input was $x$ if that $y$ was output: it is the $y$-th column (of the joint $J$)  divided by the marginal probability of that $y$, that is  $J_{-,y}/J_{\Sigma,y}$ (provided the marginal is not zero).

If we fix $\pi$ and $C$, and use the conventional abbreviation $p_{XY}$ for the resulting joint distribution $(\JD{\pi}{C})$, then the usual notations for the above are $p_X$ for left marginal $({=}\,\pi)$ and $p_X(x)$ for its value $\pi_x$ at a particular $x$, with $p_Y$ and $p_Y(y)$ similarly for the right marginal. Then $p_{X|y}(x)$ is the posterior probability of the original observation's being $x$ when $y$ has been observed. Further, we can write just $p(x)$ and $p(y)$ and $p(x|y)$ when context makes the (missing) subscripts clear.


\begin{figure}[t]
\[
  \begin{tabular}{|c|ccc|}
   \hline
   $C$     & $y_0$ & $y_1$ & $y_2$ \\ \hline
   $x_0$   & \nicefrac{2}{3} & \nicefrac{1}{6} & \nicefrac{1}{6} \\
   $x_1$   & \nicefrac{1}{3} & \nicefrac{1}{3} & \nicefrac{1}{3} \\
   $x_2$   & \nicefrac{1}{6} & \nicefrac{1}{6} & \nicefrac{2}{3} \\
   \hline
   \end{tabular} \qquad
   \Delta_C = \begin{array}{c}
    \left[
    \begin{array}{ccc}
        \nicefrac{4}{7} & \nicefrac{1}{4} & \nicefrac{1}{7} \\
        \nicefrac{2}{7} & \nicefrac{1}{2} & \nicefrac{2}{7} \\
        \nicefrac{1}{7} & \nicefrac{1}{4} & \nicefrac{4}{7}
    \end{array}
    \right] \\
    \left.
    \begin{array}{ccc}
        \nicefrac{7}{18} & \nicefrac{2}{9} & \nicefrac{7}{18}
    \end{array} 
    \right.
    \end{array}
\]
\caption{A channel $C$ and its corresponding hyper $\Delta_C$ produced by the action of the uniform distribution.}\label{fig:channel_and_hyper} 
\end{figure}


\subsection{Bayesian Adversaries and Generalised Entropy}

QIF assumes that adversaries are Bayesian: they are equipped with a prior $\pi\In\Dist{\calx}$ over secrets $\calx$ and use their knowledge of the channel $C\In\calx{\to}\Dist{\caly}$ to maximise their advantage after making an observation. This is modelled in full generality using the ``$g$-leakage framework''~\cite{m2012measuring}.
 In detail, 
 a gain function $g\In{\calw{\times}\calx}{\to}\RealNN$ models the gain to an adversary who takes an \emph{action} $w{\in}\calw$ when the value of the secret is $x{\in}\calx$. 
Focussing on actions rather than observations naturally encompasses the idea of ``remapping'' observations to their associated most likely secret value.
Moreover $g$-leakage gives a  measurement for how successful are (privacy) mechanisms at communicating useful information by comparing the ``prior''  and ``posterior'' $g$-vulnerabilities  as follows.  
 For gain function $g$ the (prior) vulnerability wrt.\ the prior $\pi$ (for a consumer/adversary with this information)  is the maximum gain over all possible actions: $V_g(\pi) \Defs \max_{w{\In}\calw} \sum_{x{\In}\calx} \pi_x g(w,x)$; it  quantifies the adversary's success in the scenario defined by $\pi$ and $g$. 
The expected \emph{(posterior) vulnerability} is the adversary's expected gain after observing $y$ is $V_g\Hyp{\pi}{C} \Defs \sum_{y{\In}\caly}  p(y) V_g(p(\calx|y))$, equivalently:  
\begin{equation}\label{e1435}
	V_g\Hyp{\pi}{C} = \sum_{y{\In}\caly} \max_{w{\In}\calw} \sum_{x\In\calx} \pi_x C_{x,y} g(w,x) ~.
\end{equation}
The greater the difference in the prior/posterior vulnerability, the better is the adversary able to use the transmitted  to infer the value of the secret. (See \Sec{ss1444} below.)

Note that we also use an equivalent formulation of leakage in terms of loss functions and  minimising ``generalised entropies''. In particular a function $\ell: \calw{\times}\calx\rightarrow \Real$ (also) defines a generalised entropy $U_\ell:\Dist\calx{\rightarrow}\Real$: 
\begin{equation}\label{e1119}
	U_\ell(\pi) \Wide\Defs \min_{w{\In}\calw} \sum_{x{\In}\calx} \pi_x \ell(w,x) ~, 
\end{equation} 
with the formulation for expected loss through a channel defined similarly using minimisation (compare \Eqn{e1435}):
\begin{equation}\label{eq:qif_loss_fn}
	U_\ell\Hyp{\pi}{C} = \sum_{y{\In}\caly} \min_{w{\In}\calw} \sum_{x\In\calx} \pi_x C_{x,y} \ell(w,x) ~.
\end{equation}

We will often use $g$ for gain and $\ell$ for loss, but either can be used in formulating a vulnerability ($V_g/V_\ell$) or generalised entropy ($U_g/U_\ell$); the leakage theory based on losses or gains is equivalent \cite{alvim2019science}.

\begin{remark}\label{rem:equivalence}
We note that the Bayesian formulation of utility loss adopted in the literature (\cite{ghosh2012universally,brenner2010impossibility}) and presented in Eqn~\ref{eq:utility_loss}, is equivalent to the formulation for posterior utility given in Eqn~\ref{eq:qif_loss_fn} (see \App{app:sec1} for proof).
\end{remark}
\noindent We adopt Eqn \Eqn{eq:qif_loss_fn} to describe utility for consumers.


Equation \Eqn{e1435}  indicates that the actual output $\caly$ can be abstracted, leaving the leakage properties of a channel in the context of a prior to be determined by a  \emph{hyper-distribution} (or simply ``hyper''). A hyper  is a distribution of distributions on $\calx$, having type $\Dist\Dist{\calx}$ or $\Dist^2{\calx}$. Given a channel $C$ and prior $\pi$, we write $\Hyp{\pi}{C}$ for the hyper whose support is the set of posterior distributions $p(\calx|y)$ on $\calx$ and which assigns the corresponding marginal $p(y)$ to each. We usually denote by $\delta^y$ the posterior $p(\calx|y)$ and by $\alpha_y$ its corresponding marginal.

An example of a channel and hyper is shown in \Fig{fig:channel_and_hyper}. The hyper $\Delta_C$ is produced by pushing the uniform prior through the channel $C$ to produce a joint distribution, which is then marginalised along its columns to produce the posteriors $\delta^y$ (the columns in $\Delta_C$) and the corresponding $y$-marginals $\alpha_y$ (the labels beneath each column). The prior $\pi$ can be recovered by averaging the posteriors $\delta_y$ via the marginals $\alpha_y$, revealing a significant correspondence between channels and hypers:
\begin{lemma}[Cor. 4.8 from \cite{alvim2019science}]\label{lem:fund_cor}
For any channel $C$ and prior $\pi$, $\pi = \sum_i \alpha_i \delta^i$ where $\Hyp{\pi}{C} = \sum_i \alpha_i[\delta^i]$~.
\end{lemma}

\Fig{fig:channel_and_hyper} illustrates this relationship between a channel and its associated hyper.

It turns out that hypers, and their refinement relation, have a compelling \emph{geometric} interpretation (first explained  in  \cite[Ch.\,12]{alvim2019science}), which we use below to study optimal utility.


\subsection{Robustness measures: refinement and channel capacities}\label{ss1444}

The study of $g$-vulnerabilities has given rise to an elegant theory of \emph{refinement}: we say that channel $A$ refines channel $B$, written $B \ChanRef A$, to mean that $A$ is safer than $B$ which holds exactly when the $g$-vulnerability of $A$ is no greater than that of $B$ for \emph{any} prior or gain function, ie., 
\begin{equation}\label{eqn:refinement}
	B \ChanRef A ~~\textit{iff}~~ V_g\Hyp{\pi}{A} \leq V_g\Hyp{\pi}{B} ~~\textit{for all}~ \pi\In\Dist{\calx}, g{\in}\Gain{\calx} 
\end{equation} 
The above holds dually for $U_\ell\Hyp{\pi}{A} \geq U_\ell\Hyp{\pi}{B}$. Remarkably, this holds whenever there exists a channel $P$ such that $BP = A$ (writing $BP$ for matrix multiplication)~\cite{McIver:2014aa}.  In fact $P$  corresponds to a ``postprocessing'' of the output of a channel, always suppressing information flow.  The channel $\mathbb{I}$ with a single column of $1$'s is maximal in the refinement order, and leaks nothing at all i.e.\ $U_\ell\Hyp{\pi}{\mathbb I}= U_\ell(\pi)$.

{Note that refinement is a strong universal utility property of channels (and therefore mechanisms modelled as channels). Refinement between two channels $A$ and $B$ means that for all priors and all gain (loss) functions channel $A$ will always have more gain (less loss) when compared to the corresponding scenario for channel $B$. We shall see therefore that when $A,B$ are members of the same privacy type then a refinement relation between them implies a universal optimality property between the two. We find, however, that in general there is no mechanism that anti-refines all mechanisms within the type, and therefore the universal optimality property can only hold for a subset of specific gain/loss functions.


Equivalently, refinement can be expressed as a relation between hypers: $\Hyp{\pi}{B} \ChanRef \Hyp{\pi}{A}$ exactly when there is a ``refining Earth Move'' (defined below) from the posteriors of $\Hyp{\pi}{B}$ to the posteriors of $\Hyp{\pi}{A}$. We recall the important correspondence between refinement of channels and refinement of hypers~\cite[Ch.\,12]{alvim2019science}: 
\begin{equation}\label{eq:hyper_correspondence}
    B \ChanRef A ~~\textrm{iff}~~ \Hyp{\pi}{B} \ChanRef \Hyp{\pi}{A} 
\end{equation}
where $\pi$ is \emph{any} full support prior. In other words, refinement of channels (on all priors) is characterised by refinement of hypers on a single (full support) prior.

In this paper we use this correspondence to characterise the space of  $\dx$-private mechanisms and those which are universally $\ell$-optimal. We write $\Delta$ etc. for general hypers on $\Dist^2{\calx}$; we note that we can depict $\Delta$ as a weighted sum of vectors, where we identify each posterior $\delta^y$ in the support of $\Delta$ as a point in $\Real^{|\calx|}$. We make the following observations \cite[Ch.\,12]{alvim2019science}):
\begin{enumerate}[a)]
\item\label{en:hyper_obsA} A ``refining Earth Move'' \cite{rachev1998mass} from $\Delta_B$ to $\Delta_A$ exists when each posterior $\delta^y$ of $\Delta_A$ is realised as a convex combination of  posteriors of $\Delta_B$ (ie. an ``Earth Move'' from $\Delta_B$ to each $\delta^y$) which exactly match the marginals of $\Delta_B$ to produce the corresponding marginals on $\Delta_A$.  

\item\label{en:hyper_obsB} \label{enum:posteriors} If $\Delta_B$'s posteriors are linearly independent (as vectors), the Earth Move from a) exists whenever the posteriors of $\Delta_A$ lie inside the convex hull of the posteriors of $\Delta_B$.   
\item The difference between the posterior  and prior vulnerabilities gives a measure of the accuracy of the channel. The additive $g$-leakage is $|V_g[\pi]-V_g\Hyp{\pi}{A}|$, and the multiplicative $g$-leakage is $V_g\Hyp{\pi}{A}/V_g[\pi]$. The corresponding channel capacities take the maximal difference over  all priors and $\underline{\Gain{\calx}}$, the $0/1$-bounded vulnerabilities.
\footnote{i.e.\ $0\leq V_g[\pi]\leq 1$}
Specifically the multiplicative and additive capacities respectively are:
\begin{equation}\label{e1248a}
{\cal ML}^\times(C) ~\Defs \max_{\pi, g} (V_g\Hyp{\pi}{C}/V_g[\pi])
\end{equation}
and
\begin{equation}\label{e1248b}
{\cal ML}^+(C) ~\Defs \max_{\pi, g} |V_g[\pi]-V_g\Hyp{\pi}{C}|~.
\end{equation}
It is easy to see that $1{\leq} {\cal ML}^\times(C) {\leq} |\calx|$ and $0{\leq} {\cal ML}^+(C) {\leq} 1$. The larger either capacity, the more accurate the channel. For example, ${\cal ML}^+({\mathbb I})= 0$, and ${\cal ML}^\times({\mathbb I})= 1$ consistent with it  transmitting no information at all.  Capacities provide an alternative robust bound on the ability of any mechanism in a privacy type to deliver accurate information since it represents a tight upper bound on the maximum leakage of any channel in the type. We show how to compute capacities for the whole type either  by proving a universal optimality result or using the characterisation given in \Sec{s1225} below.
\end{enumerate}

We shall show that irrespective of whether a type can support universal optimality results, the capacities are well-defined and there exist mechanisms that achieve the capacity. Such mechanisms represent the most efficient implementations that respect the privacy threshold whilst at the same time delivering the most information about the utility.

\subsection{Universal optimality and refinement}

Given the equivalence between utility notions in the literature and the QIF notion of posterior $\ell$-uncertainty (Remark~\ref{rem:equivalence}), there is then a natural connection between utility and refinement which extends to universal optimality. In its most general form, we define universal optimality as:

\begin{definition}{(Universal optimality)}\label{d1310a}
Given a class of mechanisms ${\cal T}$ we say that $M\in {\cal T}$ is universally optimal (wrt.\ ${\cal T}$) if $U_\ell\Hyp{\pi}{M}\leq U_\ell\Hyp{\pi}{M'}$ for all ${M'}{\in} {\cal T}$, all loss functions $\ell$ and all priors $\pi\in\Dist\calx$.
\end{definition}

Using Eqn~\ref{eqn:refinement}, this says that $M \in {\cal T}$ is universally optimal whenever $M' \ChanRef M$ for all $M' \in {\cal T}$. This definition appears to be too strong; indeed most mechanisms --even within the same privacy type-- are not related by refinement~\cite{Chatzi:2019}.  Surprisingly, we find that there is a class ${\cal T}$ for which this strong notion of universal optimality holds (\Cor{cor:1123}). However, we also find that this class is unique (\Thm{thm:fund_imp}) and universal optimality is too strong in general.

Therefore, following Ghosh et al.\ we study a weaker order defined by specific loss functions: 

\begin{definition}{(Universal $\ell$-optimality)}\label{d1310}
Given a loss function $\ell$ and a class of mechanisms ${\cal T}$ we say that $M\in {\cal T}$ is universally $\ell$-optimal (wrt.\ ${\cal T}$) if $U_\ell\Hyp{\pi}{M}\leq U_\ell\Hyp{\pi}{M'}$ for all $M'{\in} {\cal T}$ and priors $\pi\in\Dist\calx$.
\end{definition}

The universal optimality definition adopted by Ghosh et al. is, using \Def{d1310}, universal $\ell$-optimality for all ``monotone'' loss functions $\ell$ (see \Def{d1609} ahead).  We can think of \Def{d1310} as describing a type of restricted refinement which holds for all priors but only for loss functions $\ell \in {\mathcal{L}}$.  This idea of refinement restricted to loss function classes has not to our knowledge been previously studied, and is the focus of \Sec{sect:utility}. 

%% file: privacy_types.tex

In this section we study mechanisms within a privacy type, finding a new characterisation of the type requiring only a finite set of distinguished mechanisms which we call \emph{kernel mechanisms}. This characterisation captures the leakage properties of the privacy type, reducing the study of optimality to the study of kernel mechanisms.

\subsection{Privacy types as hyper-distributions}

Following \cite{DBLP:conf/icalp/AlvimACP11}, we write a mechanism $M$ of privacy type $\PrivType$ as a channel $C\In\CalX \to \Dist{\caly}$ satisfying
	$C_{x,y} \leq e^{\dx(x, x')} C_{x',y}$
for all $x, x' \in \calx$ and $y \in \caly$. ie., As a set of constraints on each column of the channel.
Importantly, there is a correspondence between $C$ and hyper-distributions as follows:
%

\begin{restatable}{lemma}{LemConstraints}\label{eq:dxconstraints}
C is $\dx$-private iff $\delta^y_x {-} e^{\dx(x, x')} \delta^y_{x'}  \leq 0$ for every $\delta^y$ in the support of $\Hyp{\Uniform}{C}$ where $\Uniform \in \Dist\calx$ is the uniform distribution.
\end{restatable}

\begin{figure}[!ht]
\centering
  \includegraphics[width=0.75\linewidth]{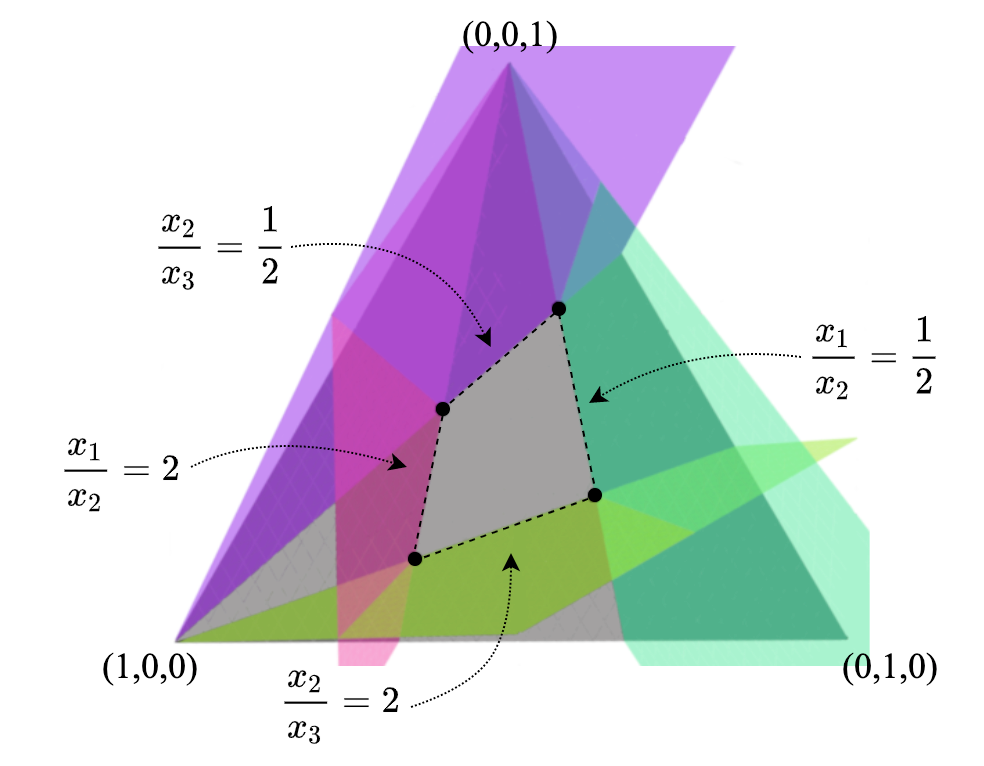}  
\caption{The space of $\MetricPrivType{\calx,\Euclid}$ hypers on $\calx = \Set{x_1, x_2, x_3}$ for  $\epsilon = \ln 2$ and $\Euclid(x_1, x_2) = 1$, $\Euclid(x_2, x_3) = 1$, $\Euclid(x_1, x_3) = 2$. The coloured hyperplanes are the linear constraints which intersect the (triangular) simplex in $\Real^3$, forming the convex region shown by the dotted lines. Every point inside this region represents a $\ln2{\cdot}\Euclid$-private posterior, and any $\ln2{\cdot}\Euclid$-mechanism (channel) corresponds to a convex sum of posteriors that average to the uniform prior $\Uniform$.}
\label{fig:constraints}
\end{figure}

Applying \Lem{eq:dxconstraints} to each pair $x, x'$ yields a feasible region of hypers (defined by their posteriors $\delta^y$) enclosed by hyperplanes --therefore convex-- which we call ``the space of $\PrivType$ hypers''.
\Fig{fig:constraints} illustrates one such space.

Observe that although \Lem{eq:dxconstraints} specifies the uniform distribution,  Eqn~\Eqn{eq:hyper_correspondence} says that this restriction does not apply to channel refinement. In other words, we can reason about channel refinement (over all priors) by reasoning in the space of hypers w.r.t.\ the uniform distribution. This differentiates our approach from standard convex optimisation formulations of the differential privacy constraints on channels~\cite{Bordenabe:14:CCS} which require quantification over all priors.


\subsection{A complete characterisation of privacy types}

The convex space above describes the complete set of potential posteriors which can be found in hyper-distributions $\Hyp{\Uniform}{C}$.  Using \Lem{lem:fund_cor}, we can construct valid channels from a set of posteriors, provided that they contain the uniform distribution and can be averaged to the uniform prior.
%
%
Moreover, \Lem{eq:dxconstraints} says that if the posteriors satisfy the $\dx$-privacy constraints, then so will the corresponding channel.  (Notice this in \Fig{fig:channel_and_hyper}: the ratio between elements in each column of $C$ matches the corresponding ratios in $\Delta_C$). This means we can construct $\dx$-private channels from sets of $\dx$-private posteriors. We use this fact now to examine optimality via hypers.


We distinguish 2 types of hypers (and their corresponding mechanisms) formed from vertices in the space of $\PrivType$ hypers:

\begin{definition}{(Vertex Mechanism/Hyper)}\label{d2653}
Let $V \in \PrivType$ be a mechanism with corresponding hyper $\Delta = \Hyp{\Uniform}{V} $. We say that $V$ is a \emph{vertex mechanism} (and $\Delta$ a \emph{vertex hyper}) if the inners in $\Supp{\Delta}$ are vertices in the space of $\PrivType$ hypers.
\end{definition}

Vertex mechanisms are of particular interest because of their refinement properties: they are the minimal elements in the refinement order in their privacy type (that is, they have no anti-refinements which satisfy the privacy constraints). Put another way, it means that every $\PrivType$ mechanism is a refinement of a vertex mechanism. (See \App{app:sec2} for details). However, (potentially) infinitely many exist in any privacy type. We will instead make use of particular vertex mechanisms that we call  \emph{kernel mechanisms}.

\begin{definition}{(Kernel Mechanism/Hyper)}\label{d4321}
Let $K \in \PrivType$ have corresponding hyper $\Delta = \Hyp{\Uniform}{K}$.
We say that $K$ is a \emph{kernel mechanism} and $\Delta$ a \emph{kernel hyper} if $K$ is a vertex mechanism, and if the inners in $\Supp{\Delta}$ are linearly independent (as vectors).
\end{definition}


Importantly, every vertex mechanism can be expressed as a convex sum of kernel mechanisms, and these generate all $\PrivType$ mechanisms. This leads to the main result of this section.

\begin{restatable}{theorem}{CorKernel}{(Fundamental Characterisation of $\PrivType$ Mechanisms)}\label{l4398}
Every $\PrivType$ mechanism is a refinement of a convex sum of kernel mechanisms.
\end{restatable}

The usefulness of this characterisation is that we can now explore questions about optimality using a finite set of kernel mechanisms as a basis for generating classes of universally optimal mechanisms.

Note that our notion of privacy types plays a role similar to that of ``privacy constraint graph'' in 
\cite{brenner2010impossibility}, but it is strictly more general. First of all, the  ``privacy (constraint) graph'' is defined only for the central model, while, as explained in the introduction, the privacy types of $\dx$-privacy subsume and generalise both the central and the local model. Furthermore, from a technical point of view,  privacy graphs can be expressed using a metric, but the vice-versa is not true, since in privacy graphs there is no notion of distance between nodes.

%% file: utility.tex
As mentioned in \Sec{s1106} we shall use loss functions ${\cal L}$ (dual to gain functions) to determine the average accuracy (utility) of a mechanism with respect to a Bayesian consumer.

\subsection{Classes of loss functions}

A loss function  defines a \emph{generalised entropy} of type $\Dist\CalX\rightarrow \Real$. Such entropies are concave (and continuous), i.e. $U_\ell(\pi_1 +_p \pi_2)\geq pU_\ell(\pi_1) +(1{-}p)U_\ell(\pi_2)$. There are many loss functions that have been identified as useful for analysing information flow properties, including accuracy of mechanisms.  

\begin{figure}[!t]
\centering

      \includegraphics[width=0.8\linewidth]{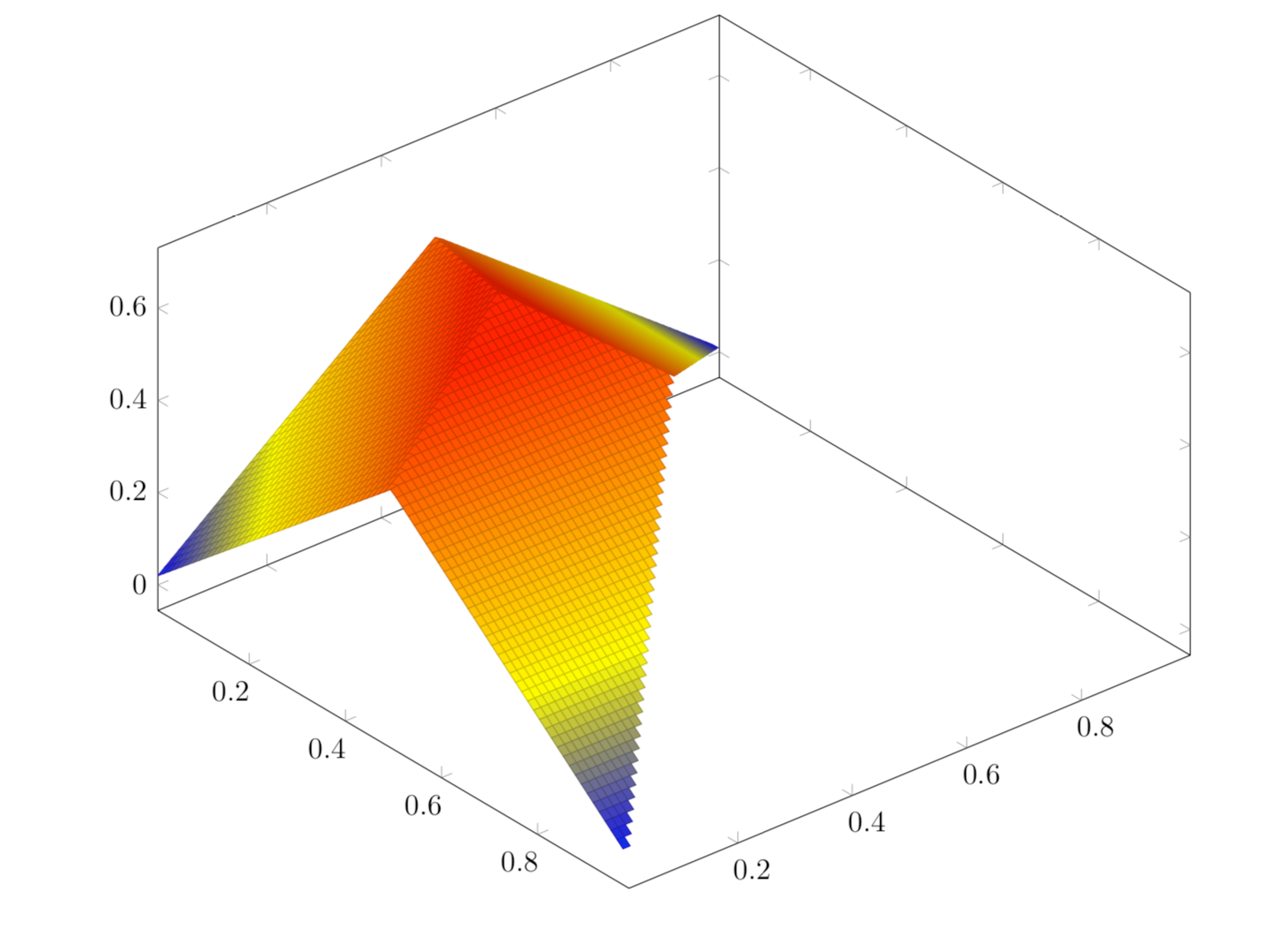} \\

\caption{$U_{\lnib}(\pi)$  for $\calx=\{x_1, x_2, x_3\}$. We can see that $\lnib$ is non-trivial because $U_{\lnib}$ is not planar.}\label{f1204}
\end{figure}

\begin{definition}[Monotone, strictly monotone and trivial loss functions]\label{d1609}
Loss functions are said to be \emph{monotone in $\dx$} denoted  ${\cal L}_\dx^m$, when they have the form  $\ell(w, x)= m(\dx(x, \alpha(w)))$, where $m$ is a monotone, non-decreasing function on the reals, and $\alpha: {\calw}\rightarrow {\calw}$ is an injective function.   The \emph{strict monotone} subclass has $m$ a strictly increasing function. The \emph{trivial} loss functions ${\cal L}^\star$ are independent of $\calw$, so have the form $f(x)$ for some real-valued function $f$. We have the following relationships between these loss functions:
\[
 {\cal L}^\star \Wide{\subseteq} {\cal L}_\dx^{m^+}\Wide{\subseteq} {\cal L}_\dx^{m} \Wide{\subseteq} {\cal L}~.
\]
\end{definition}
 Observe that the trivial loss functions are independent of the underlying metric and cannot be used to measure any non-trivial information flow property because they correspond to adversarial settings where the adversary does not have a choice of actions.
 The associated  entropies of trivial loss functions, namely $U_\ell$ correspond to linear functions in $\Dist\calx$, i.e.\ $U_\ell(p\times\pi_1 + (1{-}p)\times\pi_2)= p\times U_\ell(\pi_1)  + (1{-}p)\times U_\ell(\pi_2)$, and in fact $U_\ell \Hyp{\Uniform}{M} = U_\ell[\Uniform]$ whatever the mechanism $M$. 
Non-trivial loss functions --corresponding to non-planar generalised entropies-- are critical for measuring information flow properties as they characterise how much an adversary is able to use partial information leaks to further his intent. For example, $\lnib$ (defined in Eqn \Eqn{e1516b}) can be seen to be non-trivial as depicted in  \Fig{f1204} at right: the planar regions represent the action taken by the adversary to \emph{minimise} his loss relative to his (current) knowledge of the secret as represented by eg.\ a  posterior distribution.

Ghosh et al.\ \cite{ghosh2012universally} were the first to identify some important structures in the relation between loss functions and privacy mechanisms. They identified the class of monotone loss functions relative to the privacy type $\MetricPrivType{\mathbb{N}, \Euclid}$. In their formulation the function
$\alpha$ acts as a ``remapping'' to select the most likely input for a given observation made as part of the privacy mechanism (here represented as a channel).

Finally we note that we shall see examples later that show that the consideration of whether a loss function is included in the monotone set is sensitive to the underlying metric.

\subsection{Examples of significant loss functions}


The function $\lbin$ is a loss function commonly used to measure the ability of an adversary to guess the secret: note that its corresponding entropy $U_{\lbin}$ is referred to as \emph{Bayes' Risk}.
Letting $\calw=\calx$, we define:
\begin{equation}\label{e1516a}
\lbin(x, w) \Wide{\Defs} 0 ~~ \textit{if}~~ (x=w) ~~ \textit{else} ~~ 1~.
\end{equation}
Its dual is defined 
\begin{equation}\label{e1516b}
\lnib(x, w) \Wide{\Defs} 1 ~~ \textit{if}~~ (x=w) ~~ \textit{else} ~~ 0~.
\end{equation}
Observe that $\lbin \in {\cal L}^{m}_{\dx_2}$, but  $\lbin \not\in {\cal L}^{m^+}_{\dx_2}$, whilst $\lbin \in {\cal L}^{m^+}_{\dx_D}$.  Meanwhile $\lnib$ is not monotone in any metric, but it is non-trivial and we shall see that it is significant for computing channel capacities.

When $\calw$ and $\calx$ are subsets of a Euclidean space, we can define an average loss function:
\begin{equation}\label{e1535}
\ell_\mathrm{Avg}(x, w) \Wide{\Defs} \Euclid(w, x)~.
\end{equation}
Note that $\ell_\mathrm{Avg}$  is strictly monotone in the type $\MetricPrivType{{\mathbb N}, \Euclid}$;   the corresponding entropy $U_{\ell_\mathrm{Avg}}$ computes the average distance from the true value of the secret.

Finally we also consider
general non-monotone loss functions, where $\ell$ is an arbitrary real-valued loss function. Although these have not yet been studied for privacy mechanisms, they offer significant insights because they can express generic properties for studying property inferences \cite{DBLP:conf/concur/AlvimFMN20,alvim2019science}, and they can provide quantitative robustness measures due to the ``miracle theorems'' which quantify channel capacity. We  recall these important results as follows. 

\begin{theorem}[Multiplicative Miracle theorem \cite{alvim2019science}]\label{t1210}
The following%
\footnote{Note that multiplicative capacity is defined by maximising, hence $V_g$.}
holds for all mechanisms $M$, non-negative gain functions $g$ and priors $\pi$:
\[
V_g\Hyp{\pi}{M}/V_g[\pi] \Wide{\leq} V_{\lnib}\Hyp{\Uniform}{M}/V_{\lnib}[\Uniform]~,
\]
where
$
V_{\lnib}\Hyp{\Uniform}{M}= 1- U_{\lbin}\Hyp{\Uniform}{M}~.
$
\end{theorem}

\begin{theorem}[Additive Miracle theorem \cite{alvim2019science}]\label{t1219}
The following holds for all mechanisms $M$, 0/1-bounded vulnerabilities $V_g$ and priors $\pi$:
\[
V_g\Hyp{\pi}{M}-V_g[\pi] \Wide{\leq} 1-|\CalX|U_{\lnib}\Hyp{\Uniform}{M}~.
\]
\end{theorem}
Observe that \Thm{t1210} and \Thm{t1219} can give robust \emph{quantitative bounds} on general information leakage properties (including utility) for privacy types discussed above. We will illustrate this for the Euclidean and Discrete metrics below.

\subsection{Which mechanisms have non-trivial $\ell$-optimal loss functions?}\label{sec:dual_space}
  One way to understand the $\ell$-optimality problem within a type is to make the association of the loss function with an  adversary (or consumer as in Ghosh et al. \cite{ghosh2012universally}) explicit. Recall that the loss function formalises an adversary's intent by describing the cost of taking action $w$ when the secret is $x$. This means that a loss function is actually universally optimal for mechanism $M$ exactly when the adversary (formalised by $\ell$) can best make use of the information leaked by $M$ compared with any other mechanism in the privacy type. These adversaries prefer $M$ over all other mechanisms in the privacy type, independent of their prior knowledge.  
Thus rather than starting with a loss function $\ell$ and asking whether it has an $\ell$-optimal mechanism, instead given a mechanism $M$  ask which adversaries (loss functions) would prefer $M$ over all others. We study that idea next.

\begin{definition}[Utility set]\label{d1849}
Let $M\in \PrivType$. Define $L^{\PrivType}_M \Defs \{\ell\in {\cal L} ~| ~ M ~\textit{is universally $\ell$-optimal within~} \PrivType\}$
\end{definition}

Notice that $L^{\PrivType}_M$ is closed under addition and multiplication by a non-negative generalised scalar, where addition is defined:
\begin{equation}\label{e1231a}
(\ell_1 {+} \ell_2)(x, (w_1, w_2)) \Defs \ell_1(x, w_1) + \ell_2(x, w_2)~,
\end{equation}
and generalised scalar multiplication is defined for (non-negative) real-valued function $v$:
\begin{equation}\label{e1231b}
(v{\times}\ell)(x, w) \Wide{\Defs} v(x)\times \ell(x, w)~.
\end{equation}

Observe also that $L^{\PrivType}_M$ always contains ${\cal L}^\star$ the set of ``trivial'' loss functions, and the utility set for the trivial mechanism that leaks no information at all consists only of trivial loss functions.

More interestingly it turns out that the structure of a privacy mechanism is reflected in the structure of the corresponding utility sets. There are two important constructors for mechanisms where we see this relationship clearly. The first is the \emph{external probabilistic choice:}  for channels $M, M' \in  \PrivType$ we write $M {{_p}\oplus} M'$  
as the mechanism obtained by applying $M$ with probability $p$   and $M'$ with probability $(1{-}p)$.  This yields a channel containing the columns of $M$ scaled by $p$ together with the columns of $M'$ scaled by $(1{-}p)$, whose columns therefore clearly satisfy the same differentially private constraints as $M, M'$ and so $M {{_p}\oplus} M' \in \PrivType$.

The second is \emph{restriction} of a mechanism in $\PrivType$ 
to one in $\MetricPrivType{{\calx}', \dx}$ %
where $\calx' \subset \calx$. We write  $\Restrict{M}{X'}$ for the restriction of the mechanism $M\in \PrivType$ to operate only on the subset of secrets $X'$ obtained by removing rows of $M$ corresponding to the secrets $\calx-\calx'$. Notice that $\Restrict{M}{X'}$ retains the original metric $\dx$ because the differentially-private constraints are only evaluated on the rows corresponding to  secrets in $X'$.

Correspondingly in the space of loss functions we have the following constructions. Denote by $\Restrict{\ell}{X'}$ the restriction of the loss function $\ell$ to some subset $X'$ of its secrets, defined by $(\Restrict{\ell}{X'})(w,x) \Defs \ell(w, x)$ for $x \in X'$.  Conversely we can \emph{lift}  a loss function to a superset of secrets $\calx$, denoted $\Extend{\ell}{\calx}$; it is obtained by ``padding'' the loss function with 0's. That is, $(\Extend{\ell}{\calx})(w,x) \Defs \ell(w, x)$ when $x \in X \subseteq \calx$ and $0$ when $x \not\in X$.

We observe now that utility sets have a structure that is dual to the structure of mechanisms. 

\begin{restatable}[$\ell$-Optimal Duality]{theorem}{OpDuality}\label{t1901}

Let $\PrivType$ and $\MetricPrivType{X, \dx}$ be privacy types where $X \subseteq \calx$.
For mechanisms $M, M' \in \PrivType$,  we have the following.
\begin{enumerate}
\item ${\cal L}^\star \subseteq L^{\PrivType}_M$~,
\item $M \ChanRef M'$ implies $L^{\PrivType}_M \supseteq L^{\PrivType}_{M'}$~,
\item  $L^{\PrivType}_{\mathbb{I}}  = {\cal L}^\star$, where $\mathbb{I}$ is the trivial mechanism,
\item $L^{\PrivType}_{M\oplus_p M'} = L^{\PrivType}_{M} \cap L^{\PrivType}_{M'}$ for $0{<}p{<}1$~,
\item $\ell \in L^{\MetricPrivType{X, \dx}}_{\Restrict{M~}{~X}}$ if and only if $\Extend{\ell}{\calx} \in L^{\PrivType}_M$~.
\end{enumerate}
\end{restatable}

Notice that \Thm{t1901}(1) means that $L^{\PrivType}_M$ is always non-empty, containing at least the set of trivial loss functions. As might be expected, \Thm{t1901}(2) implies that the more information leaked through a mechanism, the more loss functions it has in its utility set, with \Thm{t1901}(3) showing that  the trivial mechanism, which always implements differential privacy (since it leaks no information at all), is only universally $\ell$-optimal for the set of trivial loss functions.  

Next, \Thm{t1901}(4) shows that when mechanisms are constructed using external probabilistic choice, the resulting utility set is the intersection of the utility sets of the components. 

Finally, \Thm{t1901}(5) provides an important method for discovering $\ell$-optimal results by studying simpler privacy types with fewer secrets, but having the same underlying metric. We make use of \Thm{t1901}(5) in \Sec{sec:examples} when we find universal optimality results in various metric spaces of interest.


A corollary to \Thm{t1901} is that the kernel mechanisms 
generate maximal utility sets.

\begin{corollary}\label{c1537}
If $M \in \PrivType$ is a universally  $\ell$-optimal mechanism then $\ell\in L_K^{\PrivType}$ for some kernel mechanism $K$ in the type.
\begin{IEEEproof}
Follows from \Thm{t1901}(2) and (4) as \Thm{l4398} implies $M$ refines a convex sum of kernels.
\end{IEEEproof}
\end{corollary}

But even more, in cases where there is a unique kernel mechanism generating the privacy type, that mechanism satisfies ``universal optimality''.

\begin{theorem}\label{t1123}
If $K\in \PrivType$ is the unique kernel mechanism then  $L_K^{\PrivType}= {\cal L}$.
\begin{IEEEproof}
From \Thm{l4398} we note that all mechanisms $M \in \PrivType$ must satisfy $K \ChanRef M$ and so by the definition of $\ChanRef$ (e.g.\ \Eqn{eqn:refinement} for loss functions) the result follows.
\end{IEEEproof}
\end{theorem}

It turns out that there is exactly one space for which a unique kernel mechanism exists, and from \Thm{t1123} this yields the following universal optimality result:

\begin{corollary}\label{cor:1123}
If $|\calx| = 2$, and for any metric $\dx$, the following mechanism (represented as a channel) is \emph{universally optimal}:
\begin{equation}\label{e1551}
  T = \begin{bmatrix}
     k{\cdot}e^{\dx} & k \\
     k & k{\cdot}e^{\dx}  \\
  \end{bmatrix}~,
\end{equation}
where $k = \nicefrac{1}{(1 + e^{\dx})}$ is a scaling factor to ensure rows sum to $1$.
\begin{IEEEproof}
The space of $\PrivType$ hypers on 2 inputs $\{x_1, x_2\}$ is an interval on the line $x_1 + x_2 = 1$ with only 2 vertices, corresponding to $x_1 = e^{\dx} x_2$ and $x_2 = e^{\dx} x_1$. Since these are linearly independent, they must be the posteriors of a kernel hyper (cf. \Def{d4321}) which is the only vertex hyper in the space. The result follows from \Thm{t1123}.
\end{IEEEproof}
\end{corollary}

The above result is significant in that it holds for all loss functions, and is non-trivial in the sense that binary mechanisms are of interest in the privacy community. (eg., the original Random Response mechanism of Warner which is used in Google's RAPPOR \cite{Erlingsson_CCS14}).

We shall see that using \emph{restriction} and \Thm{t1901} we will be able to build non-trivial loss functions contained in the utility sets for kernel mechanisms for any privacy type.


%% file: robustness.tex
As mentioned earlier, Brenner and Nissim \cite{brenner2010impossibility} studied the question of universal $\ell$-optimality specifically for types described in terms of ``privacy constraint graphs'', which for us can be described by using metrics. For example their ``sum queries'' can be considered as mechanisms in the privacy type $\MetricPrivType{\calx, \Discrete}$. Whilst their negative results apply only to $\lbin$ within their characterisation of privacy types in terms of constraint graphs, they leave open the more fundamental question about whether optimality for non-trivial loss functions exists more generally. The next theorem (partially) answers that question, showing that
non-trivial $\ell$-optimal mechanisms always exist even when $\dx$ is the Discrete metric, and as a consequence universal $\ell$-optimality is more common than expected. 

\begin{restatable}[Universal optimality existence]{theorem}{ThmOptimalityExistence}\label{t1604}
Given any privacy type $\PrivType$, if $|\CalX|{>} 1$ then there is a non-trivial monotone loss function $\ell$ and a mechanism $M\in \PrivType$ such
that $M$ is universally $\ell$-optimal.
\end{restatable}

In particular \Thm{t1604} applies even when the underlying metric is $\dx_{D}$. More significantly, once we have found basic non-trivial loss functions contained in a mechanisms utility set, we can construct more complex loss functions, also in the utility set, by using addition (Eqn~\Eqn{e1231a}) and scaling (Eqn~\Eqn{e1231b}). Moreover, depending on the underlying metric many kinds of loss function can be significant for evaluating accuracy. In fact $\lbin$ and $\lnib$ are particularly significant since, if a corresponding universally optimal mechanism exists within a type, a robust benchmark for optimal loss exists (see \Thm{t1619}).

%% file: impossibility.tex

The universal impossibility results of Brenner and Nissim were studied for the monotone loss function class wrt ``sum queries''. In this section we generalise this result to arbitrary metrics and loss functions. Importantly, we tighten the impossibility, showing that it holds only for the class of \emph{strictly} monotone loss functions. We note that our proof is more general and direct than Brenner and Nissim's result as we are able to access generic properties of the QIF framework.

We begin with a fundamental impossibility result:

\begin{restatable}[Impossibility of Universally Optimal Mechanisms]{theorem}{ThmUniversalImpossibility}\label{thm:fund_imp}
There are no universally optimal mechanisms in $\PrivType$ for any $\dx$ and for $|\calx| > 2$.
\end{restatable}

\Thm{thm:fund_imp} applies to all loss function classes, and thus is a strengthening of the Brenner and Nissim result in that it shows that the impossibility also applies to the ``counting query'' class of Ghosh et al. 

In terms of the weaker universal $\ell$-optimality, Brenner and Nissim's impossibility result for the Bayes' risk loss function $\lbin$ shows that, at least in the Discrete domain, there exist loss functions (and therefore consumers) for which no mechanisms are universally optimal. However, as we have seen, this is not the case for all loss functions, nor is it the case for all monotone loss functions.  The next result tightens the impossibility for sum queries, showing that it holds for the class of strictly monotone loss functions.

\begin{restatable}{theorem}{ThmImpossibility}\label{thm:impossibility}
Let $\ell$ be a strictly monotone loss function on $\dx$ for some metric $\dx$.
Then there are no universally $\ell$-optimal mechanisms in the privacy type $\MetricPrivType{\calx, \Discrete}$ over $n>2$ inputs.
\end{restatable}

\Thm{thm:impossibility} applies to many loss functions of interest. For example: the average loss function $\ell_\mathrm{Avg}$ is strictly monotone in $\Euclid$ although it is non-monotone in $\Discrete$; $\lbin$ is strictly monotone in $\Discrete$ but not strictly monotone in $\Euclid$. 

Curiously, \Thm{thm:impossibility} this depends on strictly monotone loss functions wrt \emph{any} metric $\dx$. To give some intuition about this result, the proof (detailed in \App{app:sec4}) shows that strict monotonicity (wrt \emph{any} $\dx$) implies that $\Restrict{\ell}{\Set{x_1, x_2}}$ is non-trivial for every $x_1 \neq x_2$, and that Kernel mechanisms in the $\Discrete$-privacy space always contain a pair $\{x_1, x_2\}$ for which optimality only holds on trivial loss functions; viz.\  the result rests on the behaviour of $\MetricPrivType{\calx, \Discrete}$ mechanisms on 2 secrets.

This suggests that consumers who wish to differentiate over all pairs of secrets will not be well-served by mechanisms of the $\MetricPrivType{\calx, \Discrete}$ type. However, consumers who wish to distinguish between a \emph{particular} pair of secrets can expect better utility, as optimal mechanisms exist in this case. (This follows from \Thm{t1604}). This is really the nature of the discrete metric -- it can distinguish between individual pairs of inputs, but not all pairs at the same time. This highlights the importance of understanding the metrics involved when designing mechanisms with the utility of consumers in mind.

Finally, the next corollary follows from the proof of \Thm{thm:impossibility} (detailed in \App{app:sec4}):

\begin{restatable}{corollary}{CorNoLbin}
There are no universally $\lbin$-optimal mechanisms of type $\MetricPrivType{\calx, \Discrete}$ for $|\calx| > 2$.
\end{restatable}

This is a strengthening of the result in \cite{brenner2010impossibility} which proves the above for $\epsilon$ over some threshold. Our result holds for all $\epsilon$.



\section{Universal privacy capacities}\label{s0856}

\Sec{impossibilities} teaches us that universal $\ell$-optimality can be difficult to obtain for arbitrary privacy types. In spite of this it would still be helpful to have some understanding about the ability of the mechanism to transmit useful information whilst still satisfying its privacy constraints. 
We introduce a weaker universal property of called \emph{privacy  type capacity} which we show is always well-defined when there are only finitely-many secrets. The privacy type capacity is an upper bound on the leakage of any mechanism in the type, and therefore provides a universal benchmark for maximal accuracy. 

\begin{definition}{(Privacy type capacity)}\label{d1026}
Given a privacy type $\PrivType$ we define multiplicative and additive \emph{privacy type capacities} as follows:
\begin{equation}\label{e1250a}
{\cal ML}^\times(\PrivType) ~\Defs \sup_{C \in \PrivType} ({\cal ML}^\times(C))
\end{equation}
and
\begin{equation}\label{e1250b} 
{\cal ML}^+(\PrivType)) ~\Defs \sup_{C \in \PrivType}  ({\cal ML}^+(C)) ~.
\end{equation}
\end{definition}

\begin{theorem}[Privacy type capacity is well defined]\label{t1038}
Given any privacy type $\PrivType$, if $|\CalX|$ is finite then both multiplicative and additive type capacities exist (\Def{d1026}) and for each, there exists  mechanisms in $\PrivType$ which achieve the capacity bound.
\begin{IEEEproof}
From the fundamental characterisation \Thm{l4398} every mechanism in $\PrivType$ is a refinement of a convex sum of Kernel mechanisms. When $|\CalX|$ is finite there are only finitely many of these. This means that the maximal capacity amongst Kernel mechanisms dominates the capacity of an convex sum of them, or refinement of them. The result now follows.
\end{IEEEproof}
\end{theorem}

The next result shows the relationship between type capacity and universal optimality results: since $\lbin$ and $\lnib$ are the gain functions that compute the capacity, if they belong to a mechanism's associated set of universal optimality gain functions then they will achieve the privacy type bound. 

\begin{theorem}[Universal optimal capacity]\label{t1619}
Given any privacy type $\PrivType$, if $\lbin \in L^{\cal T}_M$ then ${\cal ML}^\times(M) \geq {\cal ML}^\times(M')$ for any mechanism $M'\in \PrivType$. Similarly if $\lnib \in L^{\cal T}_M$ then ${\cal ML}^+(M) \geq {\cal ML}^+(M')$ for any mechanism $M'\in \PrivType$.
\end{theorem}  
%

The  proof appeals to the additive and multiplicative miracle theorems \cite{alvim2019science} (restated at \Thm{t1210} and \Thm{t1219}).  Observe that  these results give quantitative measurements of the accuracy of 
 mechanisms within a whole privacy type.

In general, though, even when a privacy type does not contain \emph{any} mechanism whose associated optimality loss function classes contain  either $\lbin$ or $\lnib$, we are still able to compute the capacity and discover a mechanism that achieves the capacity bounds using convex optimisation.


\begin{lemma}\label{l1854}
 The mechanism that achieves the optimal capacity within a privacy type $\PrivType$ can be written as a channel with no more than $|\calx|$ columns.
\end{lemma}

\begin{restatable}{corollary}{CorAddCapacity}\label{c1615}
The additive capacity for privacy type $\PrivType$ where $\calx= \{x_1, \dots x_n \}$ is $1-c$ where $c$ is the optimal value for the linear optimisation problem described by:

\[
\begin{array}{cc}
 \textit{minimise}~ m_{11}+\dots+ m_{nn}    &  \\
    \sum_j m_{ij} = 1 & 1\leq i\leq n\\
    m_{ij} \leq m_{kj }\times e^{d(x_i, x_{k})}                 & 1\leq i, k \leq n\\
    m_{ij}\geq 0 & 0\leq i, j \leq n
\end{array}
\]
\end{restatable}

Similarly we can compute the multiplicative capacity within a type.

\begin{corollary}\label{c1615b}
The multiplicative capacity for privacy type $\MetricPrivType{\calx, d}$ where $\calx= \{x_1, \dots x_n \}$ can be found as a linear optimisation problem described by:

\[
\begin{array}{cc}
 \textit{maximise}~ m_{11}+\dots+ m_{nn}    &  \\
    \sum_j m_{ij} = 1 & 1\leq i\leq n\\
    m_{ij} \leq m_{kj }\times e^{d(x_i, x_{k})}                 & 1\leq i, k \leq n\\
    m_{ij}\geq 0 & 0\leq i, j \leq n
\end{array}
\]
\end{corollary}

Observe that for both capacities the optimisation problem contains $O(|\calx|^2)$ constraints. In some situations this number can be reduced, depending on the properties of the metric. 

Note that \Cor{c1615} and \Cor{c1615b} are both examples of the optimisation formulations similar to that used by Ghosh et al. \cite{ghosh2012universally} for loss functions related to ``counting queries''. 
Significant here though is that our results produce universal measurements for privacy type capacity as well as a simplification of how to compute the average minimal loss, yielding the simple linear objective function  given here (rather than minimising over a convex function as in Eqn \Eqn{eq:utility_loss} or \Eqn{eq:qif_loss_fn} which accounts for the ``remapping'' feature).

%% file: examples.tex

In this section we focus on characterisations for some metric spaces of interest, including the two spaces addressed in the literature: $(\mathbb{N}, \Euclid)$ and $(\calx, \Discrete)$ for which the results of Ghosh et al. and Brenner and Nissim hold respectively.
We enumerate the kernel mechanisms in each of these spaces by computing the extreme points of the convex region of hypers, and then finding sets of (up to) $n$ posteriors which average to the uniform distribution.

\medskip

\subsection{\bf Privacy type defined by $(\mathbb{N}, \Euclid)$}
This privacy type is the one studied by Ghosh et al., corresponding to ``counting queries''.  Although rich with mechanisms, the Geometric mechanism is the only well-known of this type. It has two instances -- the (infinite) Geometric mechanism with outputs over the whole of $\Integer$ and the \emph{truncated} Geometric mechanism, in which the output domain matches the input domain (typically a subsequence of $\mathbb{N}$). 
These instances in fact have the same leakage properties, ie., they are \emph{equivalent} as channels and thus produce the same hyper-distribution~\cite{chatzikokolakis2021refinement}.

\begin{restatable}[Geometric mechanism]{definition}{DefGeometricMechanism}\label{def:geometric_construction}
The $\alpha$-geometric mechanism $G\In\calx \to \Dist{\Integer}$ has the following channel matrix:
\[
             G_{x,y} = \frac{1 - \alpha}{1 + \alpha} \cdot \alpha^{\Euclid(x, y)} 
\]
where $\alpha \in (0,1]$. This mechanism satisfies $\epsilon{\cdot}\Euclid$-privacy where $\Euclid$ is the Euclidean metric and $\epsilon = -\ln \alpha$.
\end{restatable}

Unsurprisingly (given its optimality properties), we find that the Geometric mechanism is a kernel mechanism.

\begin{restatable}{lemma}{LemGeomKernel}\label{l4990}
The (infinite/truncated) Geometric mechanism is a kernel mechanism of privacy type $\MetricPrivType{\mathbb{N}, \Euclid}$.
\end{restatable}

\begin{figure}[!ht]
\centering
\begin{minipage}{0.45\textwidth}
   \centering
   \small
   \begin{tabular}[t]{ c c c c c }
  \toprule 
  Dims & Vertices & \thead{Kernel \\ Mechanisms} & \thead{ Mult. \\Capacity} & \thead{ Add.\\ Capacity} \\
  \midrule
  2 & 2 & 1 & 1.33 & 0.33 \\
  3 & 4 & 2 & 1.67 & 0.5 \\
  4 & 8 & 11 & 2 & 0.67 \\
  5 & 16 & 187  & 2.33 & 0.75 \\
  6 & 32 & 15346 & 2.67 & 0.83 \\
  \bottomrule 
  \end{tabular}
  \subcaption{\small{Kernel mechanisms and capacities in the privacy type induced by ${\ln 2}{\cdot}\Euclid$.}}
  \label{tab:kernel-mechanisms} 
\end{minipage}\hspace{3mm}%
\begin{minipage}{0.45\textwidth}
 \centering
 \includegraphics[width=0.6\linewidth]{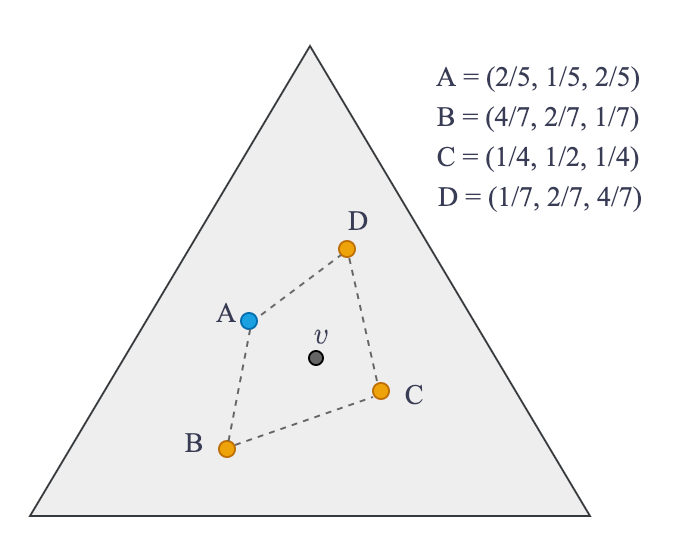} \\
 \subcaption{\small{The space of $\Euclid$-private hypers on 3 inputs $\Set{0, 1, 2}$. The Geometric mechanism for $\epsilon = \ln 2$ consists of the 3 orange vertices (B, C, D).}}
 \label{fig:GeomAsKernel}
\end{minipage}
\caption{Kernel mechanisms in the space of $\Euclid$-private hypers.}
\end{figure}

 \Fig{fig:GeomAsKernel} depicts the space of $\MetricPrivType{\mathbb{N}, \Euclid}$ mechanisms on 3 secrets. On $n$ inputs, this space is constructed from $2\times(n-1)$ linear DP constraints resulting in $2^{n-1}$ vertices -- extreme points of the feasible region of posteriors of this type.


\Tbl{tab:kernel-mechanisms} lists all kernel mechanisms for $n$ up to 6.

Ghosh et al. showed that the Geometric mechanism is universally optimal for the class of monotone loss functions, meaning that it provides the most accuracy (within its type) to Bayesian consumers described by monotone loss functions. However this result does not provide any way to determine exactly \emph{how} accurate the data release can be, even for this optimal mechanism.  A QIF analysis is able to give robust measurements for the amount of leakage as expressed by gain or loss functions; combined  with \Thm{t1619} we are now able to give robust measurements for the whole privacy class. In particular since ${\lbin}\in L_G$, we deduce that the multiplicative channel capacity of the Geometric dominates the multiplicative channel capacity for any mechanism in the whole privacy type.
Going further we can compute the exact dominating channel capacity. (Refer to Table~\ref{tab:kernel-mechanisms} for numerical examples.)

\begin{corollary}{(Dominating multiplicative capacity)}\label{c1544}
Let $N$ be the  size of $\CalX$ and let $\alpha= e^{-\epsilon}$, where $\epsilon$ is given by the definition of DP.  Then for any mechanism $M\in \MetricPrivType{\calx, \Euclid}$ we have that its multiplicative capacity is no more than $\nicefrac{(N(1{-}\alpha) + 2\alpha)}{(1{+}\alpha)}$. 
\end{corollary}

We also observe that the Geometric mechanism is the kernel mechanism which obtains the \emph{additive} capacity for the mechanisms we computed (cf. Table~\ref{tab:kernel-mechanisms}), suggesting that it may also be optimal for the non-monotone $\lnib$ loss function.

\subsection{\bf Privacy type defined by $(\calx, \Discrete)$}\label{sub:priv:discrete}
This privacy type is the one studied by Brenner and Nissim, corresponding to ``sum queries''.
For an input space $\calx$ of size $n$, the vertices of the convex region are points of intersection of $n-1$ hyperplanes, corresponding to $n-1$ distinct privacy constraints. Since there are only 2 possible distances under the discrete metric, each vertex can only contain 2 possible values, and so the number of possible vertices is given by $\sum_{i=1}^{n-1} \binom{n}{i} = 2^n - 2$.\footnote{Thinking of each vertex as having either the `high' or the `low' value, this is the number of combinations for each choice of `high' values.} We present some of these mechanisms in \Tbl{tab:discrete-mechanisms}, including the well-known ``Random Response'' mechanism.

\begin{restatable}[Random Response mechanism]{definition}{DefRandomResponseMechanism}\label{def:rr_mechanism}
The $\alpha$-random response mechanism $R\In\calx \to \Dist{\calx}$ has the following channel matrix:
\begin{align*}
             R_{x,x} &= \nicefrac{1}{k} \\
             R_{x,y} &= \nicefrac{\alpha}{k} \qquad \textrm{for } x \ne y
\end{align*}
where $k$ is a normalisation term and $\alpha \in (0,1]$. This mechanism satisfies $\epsilon{\cdot}\Discrete$-privacy where $\Discrete$ is the discrete metric and $\epsilon = -\ln \alpha$.
\end{restatable}

\begin{figure}[!ht]
\centering
\begin{minipage}{0.45\textwidth}
  \centering
  \small
  \begin{tabular}[t]{ c c c c c }
  \toprule 
  Dims & Vertices & \thead{Kernel\\ Mechanisms} & \thead{Mult. \\Capacity} & \thead{Add. \\Capacity} \\
  \midrule
  2 & 2 & 1 & 1.33 & 0.33 \\
  3 & 6 & 5 & 1.5 & 0.4 \\
  4 & 14 & 41 & 1.6 & 0.43 \\
  5 & 30 & 1291 & 1.67 & 0.44 \\
  \bottomrule 
  \end{tabular}
  \subcaption{\small Kernel mechanisms and capacities in the privacy type induced by  ${\ln 2}{\cdot}\Discrete$ .}
   \label{tab:discrete-mechanisms} 
\end{minipage}\hspace{3mm}%
\begin{minipage}{0.45\textwidth}
  \centering
 \includegraphics[width=0.65\linewidth]{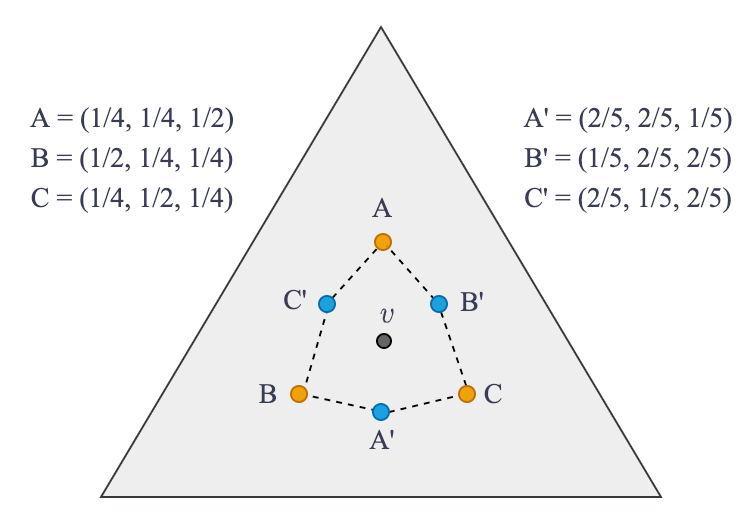} \\
   
\subcaption{\small The space of $\Discrete$-private hypers over 3 inputs. The Random Response mechanism corresponds to the 3 orange vertices (A, B, C). Its dual $R^\circ$ corresponds to the blue vertices. }
\label{fig:discrete_space}
\end{minipage}
\caption{Kernel mechanisms in the space of $\Discrete$-private hypers.}
\end{figure}


We note that the Random Response mechanism is a kernel mechanism. The space of mechanisms on 3 secrets for the Discrete metric is depicted in \Fig{fig:discrete_space}. 

\begin{restatable}{lemma}{LemRROptimality}\label{l5289}
The Random Response mechanism is a kernel mechanism in the privacy type $\MetricPrivType{\calx, \Discrete}$.
\end{restatable}


Brenner and Nissim studied mechanisms of this privacy type, focussing on Bayesian consumers and $\lbin$, showing that no mechanism is universally optimal for $\lbin$.  Their result demonstrates that there is no mechanism for the Discrete privacy type with the same universal optimality properties as the Geometric mechanism in the Euclidean privacy type. However  \Thm{t1604} indicates that there are in fact mechanisms which are $\ell$-optimal for some non-trivial (monotone) loss functions. The next example of a mechanism (also described by Brenner and Nissim) is in fact $\ell$-optimal for a non-trivial monotone loss function. 
Consider the following hypers of type $\MetricPrivType{\calx, {\ln 2}{\cdot}\Discrete}$ on $n$ inputs:

\[
    \Delta_{K_A} =
    \begin{array}{c}
    \left[
    \begin{array}{cc}
        \nicefrac{1}{4} & \nicefrac{2}{5} \\
        \vdots & \vdots \\
        \nicefrac{1}{4} & \nicefrac{2}{5} \\
        \nicefrac{1}{2} & \nicefrac{1}{5} 
    \end{array}
    \right] \\
    \left.
    \begin{array}{cc}
        \nicefrac{4}{9} & \nicefrac{5}{9}
    \end{array} 
    \right.
    \end{array}
    \hspace{0.4cm}
    \Delta_{K_B} = 
    \begin{array}{c}
    \left[
    \begin{array}{cc}
        \nicefrac{1}{2} & \nicefrac{1}{5} \\
        \nicefrac{1}{4} & \nicefrac{2}{5} \\
        \vdots & \vdots \\
        \nicefrac{1}{4} & \nicefrac{2}{5} 
    \end{array}
    \right]\\
    \left.
    \begin{array}{cc}
        \nicefrac{4}{9} & \nicefrac{5}{9}
    \end{array}
    \right.
    \end{array} 
\]

We use \Thm{t1901}(5) as follows. Concentrating on inputs $X^{\dagger} \Defs \{x_1, x_n\}$, we see that $\Restrict{K_A}{X^\dagger} {=} \Restrict{K_B}{X^\dagger} {=} T$ (where $T$ is the universally optimal mechanism from \Cor{cor:1123}). Since $T$ is the unique Kernel mechanism in a 2-element state space, we have that both are $\ell_e$-optimal for loss function $\ell_e(w, x)$ which has two actions, labelled by $X^\dagger$, and is $0$ whenever $w\neq x$, and $1$ if $w=x$. \footnote{In fact both $\Restrict{K_A}{X^\dagger},\Restrict{K_B}{X^\dagger}$ are $\ell$-optimal for all loss functions on $X^\dagger$}. 
From \Fig{f1204} we can see immediately that $\ell_e$ is non-trivial,  thus by \Thm{t1901}(5) we see that both $K_A, K_B$ are $\Extend{\ell_e}{\calx}$-optimal, as well as all probabilistic combinations $K_A \oplus_p K_B$ defined above.

While we can  compute a capacity for the whole Euclidean type from the universal optimality of the Geometric, this is not possible on the Discrete space owing to the impossibility results for $\lbin$.
We can use instead the characterisation of its structure directly to compute dominating additive and multiplicative capacities giving a tight quantitative upper bound for the accuracy of any mechanisms in the type. The dominating multiplicative capacity is given by the random response $R$, and the dominating additive capacity is given by its dual $R^\circ$ as follows. Let $\epsilon= -\ln \alpha = \ln \beta$~, and $k= 1+(|\calx|-1)e^{-\epsilon}$ and $m=1+(|\calx|-1)e^{\epsilon}$, then:

\[
\begin{array}{lll}
R_{ij} &=& 1/k~,~~\textit{if}~~ i=j\\
         & = & \alpha/k~,~~\textit{otherwise}
\end{array}
~\textit{and} ~
\begin{array}{lll}
R^\circ_{ij} &=& 1/m~,~~\textit{if}~~ i=j\\
         & = & \beta/m~,~~\textit{otherwise}~.
\end{array}
\]

The dual $R^\circ$ has the minimum elements on the diagonal whereas the Random Response $R$ has its maxima there. (Recall \Fig{fig:discrete_space} for illustration.)

\begin{corollary}[Dominating capacities]\label{c1553}
For any $M$ in the Discrete privacy type, we have ${\cal ML}^\times(M)\leq |\calx|/(1+(|\calx|-1)e^{-\epsilon})$ and ${\cal ML}^+(M)\leq 1-1/(1+(|\calx|-1)e^{\epsilon})$~.
\proof
From \Thm{l4398}, $M$ must be a refinement of a convex sum of vertex mechanisms. From the set of vertices we can see that any inner $\delta$ in the support of $\Hyp{\Uniform}{M}$ must satisfy $1/m \leq \delta_x \leq 1/k$. The result now follows from the definition of $R$ and $R^\circ$ and the miracle theorems \Thm{t1210} and \Thm{t1219}.
\end{corollary}

\subsection{\bf Privacy type defined by $(\calx, \Euclid)$ on grids}
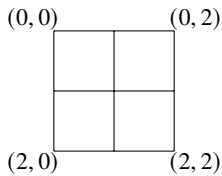
\begin{figure}[!ht]
\centering
\begin{minipage}{0.45\textwidth}
  \centering
  \small
  \begin{tabular}[t]{ c c c c c }
  \toprule 
  Dims & Vertices & \thead{Kernel\\ Mechanisms} & \thead{Mult. \\Capacity} & \thead{Add. \\Capacity} \\
  \midrule
  $1{\times}1$ & 18 & 403 & 1.68 & 0.48 \\
  $2{\times}2$ & 4798 & $>10000$ & 2.5 & 0.62 \\
  $3{\times}3$ & - & - & 3.53 & 0.79 \\
  \bottomrule 
  \end{tabular}
  \subcaption{Kernel mechanisms in the $\MetricPrivType{\calx, \Euclid}$ privacy type on grids.}
   \label{tab:grid-mechanisms} 
\end{minipage}\hspace{3mm}%
\begin{minipage}{0.45\textwidth}
 \centering
   \begin{tikzpicture}[scale=0.8,every node/.style={minimum size=.2cm-\pgflinewidth, outer sep=0pt}]
       \draw[step=1cm,color=black] (0,0) grid (2,2);
       \node (v0) at (-0.35,2.2) {\small$(0,0)$};
       \node (v1) at (2.35,2.2) {\small$(0,2)$};
       \node (v2) at (-0.35,-0.2) {\small$(2,0)$};
       \node (v3) at (2.35,-0.2) {\small$(2,2)$};
     \end{tikzpicture}
 \subcaption{The $\MetricPrivType{\calx, \Euclid}$ type on a 2x2 grid. }
  \label{fig:geolocation_grid}
\end{minipage}
\caption{The Euclidean privacy type on grids.}
\end{figure}

This privacy type is described by the Euclidean distance $\Euclid$ on $n{\times}n$ grids, as might be used in geo-location privacy, for example. Here we find that there are a large number of vertices on even the $2{\times}2$ space, making computing all of the kernel mechanisms expensive. Nevertheless, using \Cor{c1615} and \Cor{c1615b} we can efficiently compute the capacities even when the kernel mechanisms cannot be enumerated. Table \ref{tab:grid-mechanisms} illustrates some results for 3 grids.

In addition, optimality results in this space can be obtained for computed kernel mechanisms using  \Thm{t1901}(5) by employing the same technique illustrated in \Sec{sub:priv:discrete}.

\subsection{\bf Privacy type on the Hamming cube}
\begin{figure}[!ht]
\centering
\begin{minipage}{0.45\textwidth}
  \centering
  \small
  \begin{tabular}[t]{ c c c c c c  }
  \toprule 
  Bits & Dims & Vertices & \thead{Kernel\\ Mechanisms} & \thead{Mult. \\Capacity} & \thead{Add. \\Capacity} \\
  \midrule
  2 & 4 & 6 & 4 & 1.78 & 0.56 \\
  3 & 8 & 38 & $29275$ & 2.37 & 0.70 \\
  4 & 16 & - & - & 3.16 & 0.80 \\
  \bottomrule 
  \end{tabular}
  \subcaption{Kernel mechanisms in the $\MetricPrivType{\calx, \Hamming}$ privacy type for $\calx = \{0,1\}^3$.}
   \label{tab:hamming-mechanisms} 
\end{minipage}\hspace{3mm}%
\begin{minipage}{0.45\textwidth}
 \centering
   \begin{tikzpicture}[scale=2]
   	\tikzstyle{vertex}=[circle,minimum size=20pt,inner sep=0pt]
	\tikzstyle{edge} = [draw,thick,-,black]
	\tikzstyle{selected edge} = [draw,line width=2pt,-,draw_orange!80]
 	\node[vertex] (v0) at (0,0) {$000$};
	\node[vertex] (v1) at (0,1) {$001$};
	\node[vertex] (v2) at (1,0) {$010$};
	\node[vertex] (v3) at (1,1) {$011$};
	\node[vertex] (v4) at (0.23, 0.4) {$100$};
	\node[vertex] (v5) at (0.23,1.4) {$101$};
	\node[vertex] (v6) at (1.23,0.4) {$110$};
	\node[vertex] (v7) at (1.23,1.4) {$111$};
	\draw[selected edge] (v0) -- (v4) -- (v6) -- (v7);
	\draw[edge] (v0) -- (v1) -- (v3) -- (v2) -- (v0);
	\draw[edge] (v4) -- (v5) -- (v1) -- (v0);
	\draw[edge] (v7) -- (v3) -- (v2) -- (v6); 
	\draw[edge] (v7) -- (v5) -- (v4);
   \end{tikzpicture}
 \subcaption{The Hamming type on bitstrings of length 3. The orange lines depict a path on which the $\Hamming$ metric is linear. }
 \label{fig:hamming_cube_3}
\end{minipage}
\caption{The privacy type defined by the Hamming cube.}
\end{figure}
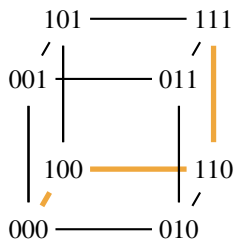

This privacy type is defined by the Hamming distance $\Hamming$ on bitstrings. As with other examples, the dimensions of this type is the number of rows in the mechanism, ie., $2^n$ for bitstrings of length $n$, and the vertices are the extreme points of the feasible region of posteriors. Here, even on bitstrings of length 3 the number of kernel mechanisms exceeds $29000$, although we again are able to compute capacities for higher dimensions. These are illustrated in Table~\ref{tab:hamming-mechanisms}.

\Fig{fig:hamming_hypers} depicts a kernel mechanism $K$  and its associated hyper $\Delta_K$ for the Hamming cube on 3 secrets. We now show how to construct a loss function $\ell$ for which $K$ is universally $\ell$-optimal.

Observe (\Fig{fig:hamming_hypers}) that for $X = \{ 000, 100, 110, 111\}$, the mechanism $\Restrict{K}{X}$ is exactly the Geometric mechanism $G$ (for $\epsilon = \ln 2$). In other words, $\lbin$ is in the utility set $L^{\MetricPrivType{\mathbb{N}, \Euclid}}_G$ which coincides with $L^{\MetricPrivType{X, \Hamming}}_G$ (noting that the Hamming metric is linear on the secrets $X$). Therefore, from \Thm{t1901}(5) we have that $\Extend{\lbin}{\calx}$ is in  $L^{\MetricPrivType{X, \Hamming}}_K$, using the fact that $\Restrict{K}{X} = G$. Therefore $K$ is universally $\Extend{\lbin}{\calx}$-optimal. And in fact $K$ will be universally $\Extend{\ell}{\calx}$-optimal for any monotone loss function $\ell$ defined on $X$.  Such a loss function would be useful to a consumer who is only interested in learning the inputs $X$ and finds no value in the remaining inputs $\calx{\setminus}X$.

Note that $X = \{ 000, 100, 110, 111\}$ corresponds to the orange path depicted in \Fig{fig:hamming_cube_3} for which we observed that the Hamming metric is linear.



\begin{figure}[!ht]
\renewcommand\arraystretch{1.2}
\[
	\Delta_K = 
	\begin{array}{c}
	\left[
	   \begin{array}{cccc}
	     \nicefrac{8}{27} & \nicefrac{2}{21} & \nicefrac{1}{21} & \nicefrac{1}{27} \\
	     \nicefrac{4}{27} & \nicefrac{4}{21} & \nicefrac{2}{21} & \nicefrac{2}{27} \\
	     \nicefrac{2}{27} & \nicefrac{2}{21} & \nicefrac{4}{21} & \nicefrac{4}{27} \\
	     \nicefrac{1}{27} & \nicefrac{1}{21} & \nicefrac{2}{21} & \nicefrac{8}{27} \\
	     \nicefrac{4}{27} & \nicefrac{4}{21} & \nicefrac{2}{21} & \nicefrac{2}{27} \\
	     \nicefrac{2}{27} & \nicefrac{2}{21} & \nicefrac{4}{21} & \nicefrac{4}{27} \\
	     \nicefrac{2}{27} & \nicefrac{2}{21} & \nicefrac{4}{21} & \nicefrac{4}{27} \\
	     \nicefrac{4}{27} & \nicefrac{4}{21} & \nicefrac{2}{21} & \nicefrac{2}{27} \\
	   \end{array}
	\right] \\
	\left.
	  \begin{array}{cccc}
	  \nicefrac{9}{32} & \nicefrac{7}{32} &  \nicefrac{7}{32} & \nicefrac{9}{32} \\
	  \end{array}
	\right.
	\end{array}
	\begin{array}{|c|cccc|}
	\hline
	K & y_1 & y_2 & y_3 & y_4 \\
	\hline
	000 & \nicefrac{2}{3} & \nicefrac{1}{6} & \nicefrac{1}{12} & \nicefrac{1}{12} \\
	100 & \nicefrac{1}{3} & \nicefrac{1}{3} & \nicefrac{1}{6} & \nicefrac{1}{6} \\
	110 & \nicefrac{1}{6} & \nicefrac{1}{6} & \nicefrac{1}{3} & \nicefrac{1}{3} \\
	111 & \nicefrac{1}{12} & \nicefrac{1}{12} & \nicefrac{1}{6} & \nicefrac{2}{3} \\
	010 & \nicefrac{1}{3} & \nicefrac{1}{3} & \nicefrac{1}{6} & \nicefrac{1}{6} \\
	011 &  \nicefrac{1}{6} & \nicefrac{1}{6} & \nicefrac{1}{3} & \nicefrac{1}{3} \\
	101 & \nicefrac{1}{6} & \nicefrac{1}{6} & \nicefrac{1}{3} & \nicefrac{1}{3} \\
	001 & \nicefrac{1}{3} & \nicefrac{1}{3} & \nicefrac{1}{6} & \nicefrac{1}{6} \\
	\hline
	\end{array}
\]
  \caption{A kernel hyper ($\Delta_K$) and corresponding kernel mechanism ($K$) on bitstrings of length $3$ wrt the Hamming distance $\Hamming$ with $\epsilon = \ln2$. $K$ is universally $\ell$-optimal for many loss functions derived from the optimality of the Geometric mechanism on $4$ secrets.}%
  \label{fig:hamming_hypers}
\end{figure}

%% file: appendix3.tex

\setcounter{theorem}{0}
\renewcommand{\thetheorem}{\Alph{section}\arabic{theorem}}

\setcounter{definition}{0}
\renewcommand{\thedefinition}{\Alph{section}\arabic{definition}}

\setcounter{lemma}{0}
\renewcommand{\thelemma}{\Alph{section}\arabic{lemma}}

\setcounter{corollary}{0}
\renewcommand{\thecorollary}{\Alph{section}\arabic{corollary}}

\subsection{Results supporting \Sec{s1106}}\label{app:sec1}

The following shows the equivalence between the formulation of utility loss in the literature (\cite{ghosh2012universally,brenner2010impossibility}) and the QIF formulation of posterior uncertainty.

\begin{lemma}\label{l3245}
The expected utility loss in Eqn~\ref{eq:utility_loss} is equivalent to the posterior $\ell$-uncertainty defined in Eqn~\ref{eq:qif_loss_fn}.
\begin{IEEEproof}
We reason as follows:
\begin{Reason}
\Step{}{
    U_\ell(\pi, M)
}
\StepR{$=$}{Eqn~\ref{eq:utility_loss}}{
   \min\limits_{r} \sum\limits_{x \in \cal{X}} \pi_x \sum\limits_{y \in \cal{Y}} M_{x,y} \; \ell(r(y), x)
}
\StepR{$=$}{Reorganising}{
   \min\limits_{r} \sum\limits_{y \in \cal{Y}} \sum\limits_{x \in \cal{X}} \pi_x  M_{x,y} \; \ell(r(y), x)
 }
 \StepR{$=$}{Min $r$ is over each $y$}{
     \sum\limits_{y \in \cal{Y}}  \min\limits_{r} \sum\limits_{x \in \cal{X}} \pi_x  M_{x,y} \; \ell(r(y), x)
 }
  \StepR{$=$}{Letting $w = r(y)$}{
     \sum\limits_{y \in \cal{Y}}  \min\limits_{w} \sum\limits_{x \in \cal{X}} \pi_x  M_{x,y} \; \ell(w, x)
 }
 \StepR{$=$}{Eqn~\ref{eq:qif_loss_fn}}{
    U_\ell[\pi{\Apply}M]
}
\end{Reason}
\end{IEEEproof}
\end{lemma}

Note that in the second-last step we employ a mapping from remaps $r(y)$ to guesses $w$. We can think of this as simply a relabelling of observations $y$ to guesses $w$, and we can always `duplicate' guesses $w$ so that there is a one-one correspondence between guesses and inputs.

\subsection{Results supporting \Sec{s1225}}\label{app:sec2}

The following establishes the correspondence between $\dx$-private mechanisms $C$ and hyper-distributions.

\LemConstraints*
\begin{IEEEproof}
Recall that $C$ is $\dx$-private whenever $\frac{C_{x,y}}{C_{x',y}} \leq e^{\dx(x, x')}$.
Now rewrite $\delta^y_x$ as $\nicefrac{\Uniform_x{\cdot} C_{x,y}}{\sum\limits_{z \in \calx} \Uniform_x {\cdot}C_{z,y}}$ and the result follows.
\end{IEEEproof}

We next set out some key properties of kernel mechanisms which allow us to establish the fundamental characterisation.
The first property tells us that kernel mechanisms are \emph{irreducible}.

\begin{restatable}{property}{PropIrreducible}\label{prop:irreducible}
If $K$ is a kernel mechanism then there is no kernel mechanism $K'$ st. $\Supp{\Delta_{K'}} \subset \Supp{\Delta_K}$ where $\Delta_K, \Delta_{K'}$ are hypers corresponding to mechanisms $K, K'$ respectively.
\end{restatable}
\begin{IEEEproof} 
Assume that $K'$ is a kernel mechanism and $K'\subset{K}$ (treating $K, K'$ as sets of posteriors). Then the posteriors of $K'$ must be linearly independent (from \Def{d4321}). Therefore the uniform distribution can be uniquely expressed as a linear combination of posteriors of $K'$. ie. We can write $u = \sum_i a_i \delta^i$ for scalars $a_i$ and posteriors $\delta^i$. But the same is true for $K$. ie. Its posteriors are linearly independent and there is a unique combination that produces the uniform distribution. ie. We write $u = \sum_j b_j \gamma^j$. But the posteriors of $K'\subset{K}$. Therefore we can write
\begin{Reason}
\Step{}{
   \sum_i a_i \delta^i
}
\StepR{$=$}{Both equal to $u$}{
   \sum_j b_j \gamma^j
}
\StepR{$=$}{$K' \subset K$}{
 \sum_i b_i \delta^i + \sum_{k = j-i} b_k \gamma^k
}
\end{Reason}
But if any of the scalars $b_k$ is non-zero, then we have 2 unique linear combinations of $\gamma^j$ which average to the uniform distribution (one involving only $a_i$ and $\delta^i$ and the other involving the $b_k$ and $\gamma^k$), contradicting the linear independence of the posteriors $\gamma^j$. Therefore we must have $b_k = 0$ and $b_i = a_i$ and so $K' = K$, contradicting $K'\subset{K}$. Thus no such $K'$ can exist.
\end{IEEEproof}

The next property says that kernel mechanisms generate all of the vertex mechanisms.

\begin{restatable}{property}{PropGenerateVertex}\label{prop:generate}
Any vertex mechanism can be written (non-uniquely) as a convex sum of kernel mechanisms.
Conversely, any convex sum of kernel mechanisms is a vertex mechanism.
\end{restatable}
\begin{IEEEproof}
For the converse, note that the convex sum of hypers takes the union of posteriors, which is clearly a vertex mechanism.
For the forward direction, we describe an algorithm whose invariant is the expected value $E$ of the remaining posteriors of $M$ (wrt the remaining outers of $M$). We start by removing probability mass corresponding to any kernel mechanism $M_1$ whose set of posteriors is a subset of the set of posteriors of $M$. Since $M_1$ is a valid mechanism, it has the same expected value $E$ as $M$, so this operation preserves $E$. We continue along this vein for every kernel mechanism we can find. Finally if we are left with some set of weighted posteriors, whose expected value must also be $E$ and so they must form a kernel mechanism, utilising the rest of the probability mass.
\end{IEEEproof}

Our final property says that kernel mechanisms are not related by refinement.

\begin{restatable}{property}{PropNoRefinement}\label{prop:no_refinement}
If $K$, $K^*$ are kernel mechanisms then $K \not\ChanRef K^*$ and $K^* \not\ChanRef K$.
\end{restatable}
\begin{IEEEproof}
Let $\Delta, \Delta^*$ be the hypers of $K, K^*$ respectively. From Property~\ref{prop:irreducible} we cannot have $\Delta\subset\Delta^*$ or $\Delta^*\subset\Delta$ (considering $\Delta, \Delta^*$ as sets of posteriors). Therefore, by convexity of the region, the posteriors of $\Delta$ cannot lie inside the convex hull of posteriors of $\Delta^*$ and vice versa. The result follows.
\end{IEEEproof}


Observe that although Property~\ref{prop:generate} says that kernel mechanisms generate the space of vertex mechanisms, the representation of a vertex hyper as a convex sum of kernel hypers is not necessarily unique.

We are now ready to prove the characterisation of $\PrivType$ mechanisms.

\CorKernel*

\begin{IEEEproof}
Given $M \in \PrivType$ which is not a vertex mechanism, and its corresponding hyper $\Delta_M$, we know that each posterior $\delta^i$ in $\Supp{\Delta_M}$ sits inside the convex hull of the vertices in the space of $\dx$-private hypers, and we can thus perform a ``reverse Earth Move'' (recall Observation \ref{en:hyper_obsA}), moving mass from each posterior $\delta^i$ to some set of vertices whose convex hull encloses $\delta^i$, preserving the overall centre of mass $\Uniform$ of $\Delta$. Thus we get a valid anti-refining vertex mechanism, which from Property~\ref{prop:generate} is a convex sum of kernel mechanisms.
\end{IEEEproof}

\subsection{Results supporting \Sec{sect:utility}}\label{app:sec4}

In this section we detail the proofs of \Thm{t1901} and \Thm{t1604} as well as the impossibility result of \Thm{thm:fund_imp} which states that there are no universally optimal mechanisms on $n>2$ inputs. We also tease out the details of the impossibility result for $\ell$-optimal mechanisms in \Thm{thm:impossibility}.

We begin with the result of \Thm{t1901} which lays out the duality between loss functions and mechanisms.

\OpDuality*

\begin{IEEEproof}
\begin{enumerate}
\item ${\cal L}^\star \subseteq L^{\PrivType}_M$~.
This follows since for $\ell \in {\cal L}^\star$ we have that $ U_\ell(\pi, M) = U_\ell(\pi)$ for any mechanism $M$. Therefore all mechanisms are $\ell$-optimal for trivial loss functions.\\

\item $M \ChanRef M'$ implies $L^{\PrivType}_M \supseteq L^{\PrivType}_{M'}$~.
Let $\ell\in L^{\PrivType}_{M'}$. We have that:
\[
U_\ell(\pi, M) \leq U_\ell(\pi, M') \leq U_\ell(\pi, M) ~, 
\]
where the first inequality holds by the refinement assumption, and the second by the universal $\ell$-optimality of $M'$ Thus $U_\ell(\pi, M) = U_\ell(\pi, M')$  and therefore $\ell\in L^{\PrivType}_{M}$.\\

\item  $L^{\PrivType}_{\mathbb{I}}  = {\cal L}^\star$, where $\mathbb{I}$ is the trivial mechanism.

By \Lem{lem:fund_cor} and \Thm{l4398} we can represent the hyperdistribution   $[\Uniform, K]$ as the sum $\sum_i a_i v_i$, where $v_i$ are vertices in the convex space. Suppose that $\ell$ is not trivial and that $U_\ell(\Uniform) = \sum_x \ell(x, w^*)$, then at least one vertex $v$ must have $U_\ell(v) = \sum_x \ell(x, w)$, where $w\neq w^*$. Assume wlog.\  that $v=v_1$. This means that:

\begin{Reason}
\Step{}{U_\ell(\Uniform, \mathbb{I})}

\Step{$=$}
{U_\ell(\Uniform)}

\Step{$=$}
{\sum_i a_i U_\ell(\Uniform)}

\Step{$=$} 
{\sum_i a_i (\sum_x \ell(x, w^*))}

\Step{$>$}
{\sum_i a_i \min_{w \in \calw}(\sum_x \ell(x, w))}

\Step{$=$}
{U_\ell(\Uniform, K)~,}
\end{Reason}
thus $\ell \not\in L^{\PrivType}_{\mathbb{I}}$.
\\

\item $L^{\PrivType}_{M\oplus_p M'} = L^{\PrivType}_{M} \cap L^{\PrivType}_{M'}$ for $0{<}p{<}1$~.
Note that $U_\ell(\pi, M\oplus_p M') = p\times U_\ell(\pi, M) + (1{-}p)\times U_\ell(\pi,M')$. Thus if $\ell \in L^{\PrivType}_{M} \cap L^{\PrivType}_{M'}$ it must also be contained in $L^{\PrivType}_{M\oplus_p M'}$.

Conversely, if $\ell \in L^{\PrivType}_{M} - L^{\PrivType}_{M'}$ then there is some $\pi$ such that $U_\ell(\pi, M) < U_\ell(\pi, M')$, in which case $U_\ell(\pi, M) < U_\ell(\pi, M\oplus_p M')$ also, since $0<p<1$.\\

\item $\ell \in L^{\MetricPrivType{X, \dx}}_{\Restrict{M~}{~X}}$ if and only if $\Extend{\ell}{\calx} \in L^{\PrivType}_M$~.
This follows since for $\pi \in {\mathbb D}X$ we have $U_\ell(\pi, \Restrict{M~}{~X}) = U_{\Extend{\ell}{\calx}}(\pi, M)$. More generally $U_{\Extend{\ell}{\calx}}(\pi, M) = \alpha \times U_\ell(\pi', \Restrict{M~}{~X})$, where $\alpha = \pi(X)$ and $\pi'_x = \pi_x/\alpha$ for $x\in X$.
\end{enumerate}
\end{IEEEproof}

Next we detail the proof of universal optimality existence, which relies on the key idea that optimality on a subset of inputs can be extended to optimality on the full mechanism.

\ThmOptimalityExistence*

\begin{IEEEproof}
Our proof is constructive - we show how to construct an $\ell$-optimal mechanism for any $\dx$ using a carefully crafted $\ell$.

For any discrete metric space $(\calx, \dx)$, choose $x_1^*, x_2^* \in \calx$ such that $\dx(x_1^*, x_2^*) \leq \dx(x_1, x_2)$ for all $x_1, x_2 \in \calx$.  Then construct the following hyper:

\[
    \begin{array}{|c | cc |}
    \hline
    \Delta & \nicefrac{k}{k+j} & \nicefrac{j}{k+j} \\
    \hline
    x_1^* & \nicefrac{1}{k} & \nicefrac{\alpha}{j} \\
    x_2^* & \nicefrac{\alpha}{k} & \nicefrac{1}{j} \\
    x_a & \nicefrac{\alpha}{k} & \nicefrac{1}{j} \\
    x_b & \nicefrac{\alpha}{k} & \nicefrac{1}{j} \\
    \ldots & \ldots & \ldots \\
    x_n & \nicefrac{\alpha}{k} & \nicefrac{1}{j} \\
    \hline
    \end{array}
\]
where $\alpha = e^{-\epsilon}$ and $k$, $j$ are normalising constants ensuring columns sum to $1$. \\
It is easy to check that the posteriors average to the uniform distribution using the outer probabilities, hence the hyper corresponds to a proper mechanism $K$. 
We now observe that $K$ is $\dx$-private for any $\dx$. For any $x_1, x_2 \neq x_1^*, x_2^*$ the $\dx$-privacy constraints are trivially satisfied in $\Delta$ and therefore in $K$ (by \Lem{eq:dxconstraints}). Also, $\delta^y_{x_1^*} \leq \delta^y_{x_2} {\times}e^\dx(x_1^*, x_2^*) \leq \delta^y_{x_2} {\times}e^\dx(x_1^*, x_2)$ (by assumption) and similarly for $x_2^*$. Therefore the $\dx$-privacy constraints are satisfied on $\Delta$ and so by \Lem{eq:dxconstraints}, $K$ is $\dx$-private.

Now, if we construct $\Restrict{K}{\Set{x_1^*, x_2^*}}$ we get:
\[
    \begin{array}{|c|cc|}
    \hline
    \Restrict{K}{\Set{x_1^*, x_2^*}} & y_1 & y_2 \\
    \hline
    x_1^* & \nicefrac{1}{(k+j)} & \nicefrac{\alpha}{(k+j)} \\
    x_2^* & \nicefrac{\alpha}{(k+j)} &  \nicefrac{1}{(k+j)}  \\
    \hline
    \end{array}
\]
which corresponds to the universally optimal mechanism on 2 inputs (\Cor{cor:1123}). Choosing any $\ell$ on 2 inputs, we can \emph{lift} it to a loss function $\Extend{\ell}{\calx}$ on the whole domain, which by \Thm{t1901}(5) is therefore universally optimal for $K$. 

\end{IEEEproof}

Finally we detail the proofs of the impossibility results.
In this section we will denote by $T$ the universally optimal mechanism on 2 secrets (cf. \Cor{cor:1123}). We will say that $K \equiv T$ whenever $K$ and $T$ represent the same abstract channel.

We first prove that for the space of $n>2$ inputs there is no single minimal element under refinement.

\begin{lemma}\label{lem:num_vertices}
Let $(X, d)$ be a metric space and let $|X| > 2$. Then the space of $d$-hypers contains at least $2(n-1)$ vertices.
\end{lemma}
\begin{IEEEproof}
Pick any 3 inputs $x_1, x_2, x_3$. If these inputs are collinear (ie. $d(x_1, x_2) + d(x_2, x_3) = d(x_1, x_3)$) then the $\dx$-privacy constraints on $(x_1, x_2)$ and $(x_2, x_3)$ imply the constraints on $(x_1, x_3)$, and so the 3 hyperplanes (corresponding to constraints in each direction) contribute only 4 vertices to the convex region. If the points are not collinear, then each pair $(x_i, x_j)$ contributes 2 constraints, resulting in 6 constraint hyperplanes in total and 6 points in the convex space. The minimum number of vertices in the space therefore occurs when the set $X$ is totally ordered (ie. maximising the number of linear, and thus inferred, constraints), and so we need only consider constraints on `adjacent' vertices. In total this yields $2(n-1)$ vertices (corresponding to 2 constraints per adjacent pair).
\end{IEEEproof}

We now have the details in place to prove the impossibility of universally optimal mechanisms for $n > 2$.

\ThmUniversalImpossibility*

\begin{IEEEproof}
By \Lem{lem:num_vertices} we have that the space contains at least $2(n-1)$ vertices.
Since this is larger than $n$ for $n > 2$, we must have more than one kernel mechanism, and thus more than one vertex mechanism. The result follows.
\end{IEEEproof}


To prove \Thm{thm:impossibility} we need a series of technical lemmas. We will first show that strictly monotone $\ell$ implies its restriction to pairs of secrets is non-trivial (\Lem{lem:non_trivial}). We then show that the latter condition implies the impossibility (\Lem{lem:impossibility1}). 

We begin by observing that trivial loss functions can be defined as follows:
\begin{definition}[Trivial Loss Function]\label{app:def:trivial}
A loss function $\ell:\calw \times \calx \Fun \RealNN$ is called \emph{trivial} if there exists a $w^* \in \calw$ such that $\ell(w^*, x) \le \ell(w, x)$ for all $x \in \calx$ and $w \in \calw \setminus \Set{w^*}$.
\end{definition}

We now have the following property of strictly monotone loss functions.

\begin{lemma}\label{lem:non_trivial}
Given a loss function $\ell:\calw \times \calx \Fun \RealNN$, if there exists a metric $\dx$ such that $\ell$ is strictly monotone in $\dx$, then $\Restrict{\ell}{\Set{x, x'}}$ must be non-trivial for all $x, x' \in \CalX$ with $x \neq x'$.
\end{lemma}
\begin{IEEEproof}
Assume that there exists $x, x' \in \calx$ with $x \neq x'$ st. $\Restrict{\ell}{\Set{x, x'}}$ is trivial. Then by \Def{app:def:trivial} there exists a $w^*$ st. $\ell(w^*, x) \le \ell(w, x)$ and $\ell(w^*, x') \le \ell(w, x')$ for all $w \neq w^*$. Since $\ell$ is strictly monotone we must have $\dx(\alpha(w^*), x) \leq \dx(\alpha(w), x)$ and $\dx(\alpha(w^*), x') \leq \dx(\alpha(w), x')$ for all $w \neq w^*$. But we also know from strict monotonicity that there exists $w_a, w_b \in \calw$ with $w_a \neq w_b$ st. $\alpha(w_a) = x$ and $\alpha(w_b) = x'$. Therefore we have
\begin{align*}
    \dx(\alpha(w^*), x) ~&\leq~ \dx(\alpha(w_a), x) ~=~ 0 \\
    \dx(\alpha(w^*), x') ~&\leq~ \dx(\alpha(w_b), x') ~=~ 0
\end{align*}
Since $\alpha$ is injective, this implies that $w_a = w_b$ and hence $x = x'$, contradicting our assumption that $x \neq x'$. Therefore no such $x, x'$ exists.
\end{IEEEproof}

Next we need the following existence theorem and its corollary which applies to the $\Discrete$ space.

\begin{theorem}\label{thm:mono_optimal}
In the space of $\Discrete$-private hypers there exist non-trivial monotone (on $\Discrete$) loss functions $\CalL$ and mechanisms which are universally $\CalL$-optimal.
\end{theorem}
\begin{IEEEproof}
The proof uses the same construction as for \Thm{t1604}, which we note must be a $\Discrete$-private mechanism.
We choose the following loss function, a lifting of an optimal loss function on 2 inputs:
\[
    \begin{array}{| c | c c |}
    \hline
    \ell & w_1 & w_2 \\
    \hline
    x_1 & 0 & 1 \\
    x_2 & 1 & 0 \\
    x_3 & 0 & 0 \\
    x_4 & 0 & 0 \\
    \ldots & \ldots & \ldots \\
    x_n & 0 & 0 \\
    \hline
    \end{array}
\]
Since $\Restrict{\ell}{\Set{x_1, x_2}}$ is monotone on $\Discrete$, we have that $\ell$ is monotone on $\Discrete$ and the mechanism $K$ (with corresponding hyper $\Delta$) is universally $\ell$-optimal. \\
\end{IEEEproof}

The following corollary explains why we can always find universally optimal mechanisms in this space.

\begin{corollary}\label{cor:optimality_disc_2secrets}
Given any pair of inputs $x_1, x_2 \in \calx$, there are always $\Discrete$-private mechanisms $M\In\calx \to \Dist{\caly}$ for which $\Restrict{M}{\Set{x_1, x_2}} \equiv T$.
\end{corollary}
\begin{IEEEproof}
Follows from the construction given in the proof of \Thm{thm:mono_optimal}.
\end{IEEEproof}

And now we show that non-triviality is sufficient for impossibility.

\begin{lemma}\label{lem:impossibility1}
There are no universally $\ell$-optimal $\Discrete$-private mechanisms over $n>2$ inputs for any loss function $\ell$ for which $\Restrict{\ell}{\Set{x, x'}}$ is non-trivial for all $x \neq x'$.
\end{lemma}
\begin{IEEEproof}
By \Cor{cor:optimality_disc_2secrets}, it is sufficient to show that for every kernel mechanism $K$ there exists some pair of inputs $x_1, x_2$ such that $\Restrict{K}{\Set{x_1, x_2}} \not\equiv T$.
Let $K$ be any $\Discrete$-private kernel mechanism on $n>2$ inputs with corresponding kernel hyper $\Delta_K$. Then $\Supp{\Delta_K}$ has at least 2 posteriors, call them $\delta^1$ and $\delta^2$. Moreover, each $\delta^i$ is a vertex and so the $\Discrete$-privacy constraints hold tightly on $n-1$ input pairs. Pick any 3 inputs $x_1, x_2, x_3$ such that $\delta^1_{x_1} =  \alpha \delta^1_{x_2}$ and $\delta^2_{x_2} = \alpha \delta^2_{x_1}$ (ie. the constraints hold in opposite directions).  This means that $\Restrict{K}{\Set{x_1, x_2}} \equiv T$. We now show that this cannot be true for both $\Set{x_1, x_3}$ and $\Set{x_2, x_3}$.
The 4 possible constructions for $\delta^1$ and $\delta^2$ restricted to $\Set{x_1, x_2, x_3}$ are:
\[
    \begin{array}{|c|cc|}
        \hline
         \Delta_1& \delta^1 & \delta^2 \\
        \hline
        x_1 & \nicefrac{\alpha}{k} & \nicefrac{1}{j} \\
        x_2 & \nicefrac{1}{k} & \nicefrac{\alpha}{j} \\
        x_3 & \nicefrac{1}{k} & \nicefrac{\alpha}{j} \\
        \hline
    \end{array}
    \hspace{5mm}
    \begin{array}{|c|cc|}
        \hline
        \Delta_2 & \delta^1 & \delta^2 \\
        \hline
        x_1 & \nicefrac{\alpha}{k} & \nicefrac{1}{j} \\
        x_2 & \nicefrac{1}{k} & \nicefrac{\alpha}{j} \\
        x_3 & \nicefrac{1}{k} & \nicefrac{1}{j} \\
        \hline
    \end{array}
\]
\[
\begin{array}{|c|cc|}
        \hline
        \Delta_3 & \delta^1 & \delta^2 \\
        \hline
        x_1 & \nicefrac{\alpha}{k} & \nicefrac{1}{j} \\
        x_2 & \nicefrac{1}{k} & \nicefrac{\alpha}{j} \\
        x_3 & \nicefrac{\alpha}{k} & \nicefrac{\alpha}{j} \\
        \hline
    \end{array}
    \hspace{5mm}
\begin{array}{|c|cc|}
        \hline
         \Delta_4 & \delta^1 & \delta^2 \\
        \hline
        x_1 & \nicefrac{\alpha}{k} & \nicefrac{1}{j} \\
        x_2 & \nicefrac{1}{k} & \nicefrac{\alpha}{j} \\
        x_3 & \nicefrac{\alpha}{k} & \nicefrac{1}{j} \\
        \hline
    \end{array}
\]
It is clear that for each possible construction of $K$ there exists a pair $x_i, x_j$ st. $\Restrict{K}{\Set{x_i, x_j}} \not\equiv T$. Therefore there is no $K$ which can be universally $\ell$-optimal for any pairwise non-trivial loss function $\ell$ and thus no universally $\ell$-optimal $\Discrete$-mechanisms exist over the space of $n>2$ inputs.
\end{IEEEproof}

We now have the following:

\ThmImpossibility*

\begin{IEEEproof}
Follows from \Lem{lem:non_trivial} and \Lem{lem:impossibility1}.
\end{IEEEproof}

\noindent And the final corollary of this section which relies on an intermediate result above:

\CorNoLbin*

\begin{IEEEproof}
This follows from \Lem{lem:impossibility1}, noting that $\Restrict{\lbin}{\Set{x, x'}}$ is non-trivial for all $x \neq x'$.
\end{IEEEproof}

\subsection{Results supporting \Sec{s0856}}\label{app:sec5}

\CorAddCapacity*

\begin{IEEEproof}
Note that \Thm{t1219} and \Def{d1026} says that the additive capacity for a mechanism is $1$ minus the sum of the column minima.  Observe that the constraint set describes mechanisms satisfying a particular privacy type. The result follows provided that we an show that there is such a mechanism whose additive capacity can be computed from the given objective function, i.e.\ $1$  minus the sum of the diagonals. From \Thm{t1038} we know that there exists a mechanism that optimises the capacity, and from \Def{d1026} it can be computed from the sum of the column minima. Note that re-ordering of columns is equivalent to refinement, and that refinement preserves membership in a privacy type. Thus if $M$ is an instance of a mechanism exhibiting optimal additive capacity within a given type, we first produce a refinement $M'$ by summing together any columns labelled $y, y'$ for which the column minima $\min_i M_{iy}$ and $\min_i M_{iy'}$ occur for the same $i$, and noting that the additive capacity of $M'$ is the same as for $M$, with $M'$ possible having fewer columns. However we note that $M'$'s column minima all occur for \emph{different} values of $i$. We then re-order the columns of $M'$, possibly adding in non-negative columns so that the column minima of $M''$ is equal to the sum of the diagonals. The result follows since $M''$'s additive capacity is the same as that of $M$ and is therefore optimal for the whole privacy type.  
\end{IEEEproof}


\subsection{Results supporting \Sec{sec:examples}}\label{app:sec6}

\LemGeomKernel*

\begin{IEEEproof}
Denote by $G, G_t$ the geometric and truncated geometric mechanisms respectively. 
The $\dx$-privacy constraints on both $G$ and $G_t$ hold with equality (by construction) and therefore on the posteriors of $\Hyp{\Uniform}{G}$ and $\Hyp{\Uniform}{G_t}$ (by \Lem{eq:dxconstraints}) -- which are the same hyper, call it $\Delta_G$~\cite{chatzikokolakis2021refinement}. We have now that $\Delta_G$ is a vertex hyper. 
Since $G_t$ is invertible~\citep{chatzikokolakis2021refinement} its columns are linearly independent and therefore so are the posteriors of $\Hyp{\Uniform}{G_t} = \Delta_G$ (by linearity). And so $\Delta_G$ is a kernel hyper corresponding to both $G$ and $G_t$.
\end{IEEEproof}

\LemRROptimality*

\begin{IEEEproof}
It is easy to see (by construction) that $R$ is $\Discrete$-private, since every column has either $R_{x,y} = R_{x',y}$ or $R_{x,y} = \alpha R_{x',y}$ for any $x, x' \in \calx$. To show it is a kernel mechanism we need to show that the hyper $\Delta_R = \Hyp{\Uniform}{R}$ has linearly independent inners, their convex hull contains the uniform distribution, and they are vertices in the space of $\Discrete$-private hypers. Observe that $R$ is doubly stochastic and thus the inners of $\Delta_R$ are exactly the columns of $R$. Observe also that each column of $R$ has $n-1$ constraints holding tightly, and thus we have that $R$ is a vertex mechanism. Now writing $R$ as 
\[
	R = \frac{1}{k} ~ \left(
		\begin{array}{ ccccc }
		\alpha & 1 & 1 & \ldots & 1 \\
		1 & \alpha & 1 & \ldots & 1 \\
		1 & 1 & \alpha & \ldots & 1 \\
		\multicolumn{5}{c}{\ldots} \\
		1 & 1 & 1 & \ldots & \alpha
		\end{array}
	\right)
\]
we can perform basic row operations, subtracting row $k-1$ from row $k$ for rows $n$ down to $2$ to yield:
\[
	R' = \frac{1}{k} ~\left(
		\begin{array}{ ccccc }
		\alpha & 1 & 1 & \ldots & 1 \\
		1 - \alpha & \alpha - 1 & 0 & \ldots & 0 \\
		0 & 1-\alpha & \alpha - 1 & \ldots & 0 \\
		\multicolumn{5}{c}{\ldots} \\
		0 & 0 & 0 & \ldots & \alpha - 1
		\end{array}
	\right)
\]	
Noting that det $R$ = det $R'$, we compute:
\begin{align*}
	\text{det } R' &~=~ \frac{1}{k^n}  \left( \alpha (\alpha - 1)^{n-1} - 1(1 - \alpha)(\alpha-1)^{n-2} + 1(1-\alpha)^2(\alpha-1)^{n-3} + \right. \\
			& \left. \qquad \qquad \ldots + (-1)^{n-1}(1-\alpha)^{n-1} \right) \\
			& ~=~ \frac{1}{k^n}  \left( (\alpha (\alpha - 1)^{n-1} + (\alpha-1)^{n-1} + (\alpha-1)^{n-1} + \ldots + (\alpha-1)^{n-1} \right)\\
			& ~=~ \frac{1}{k^n}  (\alpha - 1)^{n-1} (\alpha + (n-1))
\end{align*}
which is non-zero for $\alpha \in (0,1)$. Thus, except for the case where $\alpha = 1$ we have that the randomised response matrix is invertible, and so its columns are linearly independent. And therefore the inners of $\Delta_R =  \Hyp{\Uniform}{R}$ are likewise linearly independent.
Finally, it is easy to check that we can write the uniform distribution as the following convex combination of columns of $R$:
\[
	\frac{1}{n} R_{(-,1)} + \frac{1}{n} R_{(-, 2)} + \ldots + \frac{1}{n} R_{(-,n)}
\]
where $R_{(-, i)}$ denotes the $i$th column of $R$. Therefore this also holds for the inners of $\Delta_R$. Thus $R$ is a kernel mechanism as required.
\end{IEEEproof}

%% file: main.bbl
\begin{thebibliography}{10}
\providecommand{\url}[1]{#1}
\csname url@samestyle\endcsname
\providecommand{\newblock}{\relax}
\providecommand{\bibinfo}[2]{#2}
\providecommand{\BIBentrySTDinterwordspacing}{\spaceskip=0pt\relax}
\providecommand{\BIBentryALTinterwordstretchfactor}{4}
\providecommand{\BIBentryALTinterwordspacing}{\spaceskip=\fontdimen2\font plus
\BIBentryALTinterwordstretchfactor\fontdimen3\font minus
  \fontdimen4\font\relax}
\providecommand{\BIBforeignlanguage}[2]{{%
\expandafter\ifx\csname l@#1\endcsname\relax
\typeout{** WARNING: IEEEtran.bst: No hyphenation pattern has been}%
\typeout{** loaded for the language `#1'. Using the pattern for}%
\typeout{** the default language instead.}%
\else
\language=\csname l@#1\endcsname
\fi
#2}}
\providecommand{\BIBdecl}{\relax}
\BIBdecl

\bibitem{Dwork:06:TCC}
C.~Dwork, F.~Mcsherry, K.~Nissim, and A.~Smith, ``Calibrating noise to
  sensitivity in private data analysis,'' in \emph{In Proceedings of the Third
  Theory of Cryptography Conference (TCC)}, ser. Lecture Notes in Computer
  Science, S.~Halevi and T.~Rabin, Eds., vol. 3876.\hskip 1em plus 0.5em minus
  0.4em\relax Springer, 2006, pp. 265--284.

\bibitem{Dwork:06:ICALP}
\BIBentryALTinterwordspacing
C.~Dwork, ``Differential privacy,'' in \emph{33rd International Colloquium on
  Automata, Languages and Programming (ICALP 2006)}, ser. Lecture Notes in
  Computer Science, M.~Bugliesi, B.~Preneel, V.~Sassone, and I.~Wegener, Eds.,
  vol. 4052.\hskip 1em plus 0.5em minus 0.4em\relax Springer, 2006, pp. 1--12.
  [Online]. Available: \url{http://dx.doi.org/10.1007/11787006_1}
\BIBentrySTDinterwordspacing

\bibitem{alvim2019science}
M.~S. Alvim, K.~Chatzikokolakis, A.~McIver, C.~Morgan, C.~Palamidessi, and
  G.~Smith, ``The science of quantitative information flow,'' 2019.

\bibitem{ghosh2012universally}
A.~Ghosh, T.~Roughgarden, and M.~Sundararajan, ``Universally utility-maximizing
  privacy mechanisms,'' \emph{SIAM Journal on Computing}, vol.~41, no.~6, pp.
  1673--1693, 2012.

\bibitem{brenner2010impossibility}
H.~Brenner and K.~Nissim, ``Impossibility of differentially private universally
  optimal mechanisms,'' in \emph{2010 IEEE 51st Annual Symposium on Foundations
  of Computer Science}.\hskip 1em plus 0.5em minus 0.4em\relax IEEE, 2010, pp.
  71--80.

\bibitem{Duchi:13:FOCS}
\BIBentryALTinterwordspacing
J.~C. Duchi, M.~I. Jordan, and M.~J. Wainwright, ``Local privacy and
  statistical minimax rates,'' in \emph{Proceedings of the 54th Annual {IEEE}
  Symposium on Foundations of Computer Science (FOCS)}.\hskip 1em plus 0.5em
  minus 0.4em\relax {IEEE} Computer Society, 2013, pp. 429--438. [Online].
  Available: \url{https://doi.org/10.1109/FOCS.2013.53}
\BIBentrySTDinterwordspacing

\bibitem{chatzikokolakis2013broadening}
K.~Chatzikokolakis, M.~E. Andr{\'e}s, N.~E. Bordenabe, and C.~Palamidessi,
  ``Broadening the scope of differential privacy using metrics,'' in
  \emph{International Symposium on Privacy Enhancing Technologies
  Symposium}.\hskip 1em plus 0.5em minus 0.4em\relax Springer, 2013, pp.
  82--102.

\bibitem{andres2013geo}
M.~E. Andr{\'e}s, N.~E. Bordenabe, K.~Chatzikokolakis, and C.~Palamidessi,
  ``Geo-indistinguishability: Differential privacy for location-based
  systems,'' in \emph{Proceedings of the 2013 ACM SIGSAC conference on Computer
  \& communications security}, 2013, pp. 901--914.

\bibitem{DBLP:conf/post/FernandesDM19}
\BIBentryALTinterwordspacing
N.~Fernandes, M.~Dras, and A.~McIver, ``Generalised differential privacy for
  text document processing,'' in \emph{Principles of Security and Trust - 8th
  International Conference, {POST} 2019, Held as Part of the European Joint
  Conferences on Theory and Practice of Software, {ETAPS} 2019, Prague, Czech
  Republic, April 6-11, 2019, Proceedings}, 2019, pp. 123--148. [Online].
  Available: \url{https://doi.org/10.1007/978-3-030-17138-4\_6}
\BIBentrySTDinterwordspacing

\bibitem{fernandes2021laplace}
\BIBentryALTinterwordspacing
N.~Fernandes, A.~McIver, and C.~Morgan, ``The laplace mechanism has optimal
  utility for differential privacy over continuous queries,'' in \emph{36th
  Annual {ACM/IEEE} Symposium on Logic in Computer Science, {LICS} 2021, Rome,
  Italy, June 29 - July 2, 2021}.\hskip 1em plus 0.5em minus 0.4em\relax
  {IEEE}, 2021, pp. 1--12. [Online]. Available:
  \url{https://doi.org/10.1109/LICS52264.2021.9470718}
\BIBentrySTDinterwordspacing

\bibitem{gupte2010universally}
M.~Gupte and M.~Sundararajan, ``Universally optimal privacy mechanisms for
  minimax agents,'' in \emph{Proceedings of the twenty-ninth ACM
  SIGMOD-SIGACT-SIGART symposium on Principles of database systems}, 2010, pp.
  135--146.

\bibitem{acharya2020context}
J.~Acharya, K.~Bonawitz, P.~Kairouz, D.~Ramage, and Z.~Sun, ``Context aware
  local differential privacy,'' in \emph{International Conference on Machine
  Learning}.\hskip 1em plus 0.5em minus 0.4em\relax PMLR, 2020, pp. 52--62.

\bibitem{kairouz2016extremal}
P.~Kairouz, S.~Oh, and P.~Viswanath, ``Extremal mechanisms for local
  differential privacy,'' \emph{The Journal of Machine Learning Research},
  vol.~17, no.~1, pp. 492--542, 2016.

\bibitem{koufogiannis2015optimality}
F.~Koufogiannis, S.~Han, and G.~J. Pappas, ``Optimality of the laplace
  mechanism in differential privacy,'' \emph{arXiv preprint arXiv:1504.00065},
  2015.

\bibitem{Asi:20}
H.~A. J.~C. Duchi, ``Near instance-optimality in differential privacy,'' 2020,
  arXiv:2005.10630v1, 2020.

\bibitem{m2012measuring}
M.~S. Alvim, K.~Chatzikokolakis, C.~Palamidessi, and G.~Smith, ``Measuring
  information leakage using generalized gain functions,'' in \emph{2012 IEEE
  25th Computer Security Foundations Symposium}.\hskip 1em plus 0.5em minus
  0.4em\relax IEEE, 2012, pp. 265--279.

\bibitem{McIver:2014aa}
A.~McIver, C.~Morgan, G.~Smith, B.~Espinoza, and L.~Meinicke, ``Abstract
  channels and their robust information-leakage ordering,'' in \emph{Principles
  of Security and Trust - Third International Conference, {POST} 2014, Held as
  Part of the European Joint Conferences on Theory and Practice of Software,
  {ETAPS} 2014, Grenoble, France, April 5-13, 2014, Proceedings}, ser. Lecture
  Notes in Computer Science, M.~Abadi and S.~Kremer, Eds., vol. 8414.\hskip 1em
  plus 0.5em minus 0.4em\relax Springer, 2014, pp. 83--102.

\bibitem{rachev1998mass}
S.~T. Rachev and L.~R{\"u}schendorf, \emph{Mass Transportation Problems: Volume
  I: Theory}.\hskip 1em plus 0.5em minus 0.4em\relax Springer Science \&
  Business Media, 1998, vol.~1.

\bibitem{Chatzi:2019}
K.~Chatzikokolakis, N.~Fernandes, and C.~Palamidessi, ``Comparing systems:
  Max-case refinement orders and application to differential privacy,'' in
  \emph{Proc. CSF}.\hskip 1em plus 0.5em minus 0.4em\relax IEEE Press, 2019.

\bibitem{DBLP:conf/icalp/AlvimACP11}
\BIBentryALTinterwordspacing
M.~S. Alvim, M.~E. Andr{\'{e}}s, K.~Chatzikokolakis, and C.~Palamidessi, ``On
  the relation between differential privacy and quantitative information
  flow,'' in \emph{Automata, Languages and Programming - 38th International
  Colloquium, {ICALP} 2011, Zurich, Switzerland, July 4-8, 2011, Proceedings,
  Part {II}}, 2011, pp. 60--76. [Online]. Available:
  \url{https://doi.org/10.1007/978-3-642-22012-8\_4}
\BIBentrySTDinterwordspacing

\bibitem{Bordenabe:14:CCS}
N.~E. Bordenabe, K.~Chatzikokolakis, and C.~Palamidessi, ``Optimal
  geo-indistinguishable mechanisms for location privacy,'' in \emph{Proc.
  {CCS}}, 2014, pp. 251--262.

\bibitem{DBLP:conf/concur/AlvimFMN20}
\BIBentryALTinterwordspacing
M.~S. Alvim, N.~Fernandes, A.~McIver, and G.~H. Nunes, ``On privacy and
  accuracy in data releases (invited paper),'' in \emph{31st International
  Conference on Concurrency Theory, {CONCUR} 2020, September 1-4, 2020, Vienna,
  Austria (Virtual Conference)}, ser. LIPIcs, I.~Konnov and L.~Kov{\'{a}}cs,
  Eds., vol. 171.\hskip 1em plus 0.5em minus 0.4em\relax Schloss Dagstuhl -
  Leibniz-Zentrum f{\"{u}}r Informatik, 2020, pp. 1:1--1:18. [Online].
  Available: \url{https://doi.org/10.4230/LIPIcs.CONCUR.2020.1}
\BIBentrySTDinterwordspacing

\bibitem{Erlingsson_CCS14}
\'{U}lfar Erlingsson, V.~Pihur, and A.~Korolova, ``{RAPPOR}: Randomized
  aggregatable privacy-preserving ordinal response,'' in \emph{Proc. {CCS}},
  2014, pp. 1054--1067.

\bibitem{chatzikokolakis2021refinement}
K.~Chatzikokolakis, N.~Fernandes, and C.~Palamidessi, ``Refinement orders for
  quantitative information flow and differential privacy,'' \emph{Journal of
  Cybersecurity and Privacy}, vol.~1, no.~1, pp. 40--77, 2021.

\end{thebibliography}
